\documentclass[review,12pt,sort&compress]{elsarticle}

\usepackage{amssymb}

\usepackage{color} 
\usepackage{float} 
\usepackage{subfig}
\usepackage{booktabs}
\usepackage{tabularx} 
\usepackage{subfiles} 
\usepackage{multirow}
\usepackage{hyperref}
\usepackage{enumitem}
\usepackage{diagbox}
\usepackage{makecell}

\usepackage{dcolumn}
\usepackage{amsmath}
\usepackage{mathptmx}
\usepackage{etoolbox}

\usepackage{algorithm2e}
\usepackage{algorithmic}

\usepackage{epstopdf}

\journal{Journal of Non-Newtonian Fluid Mechanics}

\begin{document}

\begin{frontmatter}



\title{A robust numerical strategy for finding surface waves in flows of non-Newtonian liquids
\tnoteref{label_note_copyright} \tnoteref{label_note_doi}
}

\tnotetext[label_note_copyright]{\copyright 2023. This manuscript version is made available under the CC-BY-NC-ND 4.0 license http://creativecommons.org/licenses/by-nc-nd/4.0/}

\tnotetext[label_note_doi]{Accepted Manuscript for the Journal of Non-Newtonian Fluid Mechanics, v.322, 105153, 2023, DOI: 10.1016/j.jnnfm.2023.105153}


\author[label_erick]{Bruno P. Chimetta}
\author[label_erick]{Erick M. Franklin\corref{cor1}}
\ead{erick.franklin@unicamp.br}
\cortext[cor1]{Corresponding author. phone: +55 19 35213375}

\affiliation[label_erick]{organization={School of Mechanical Engineering, University of Campinas - UNICAMP},
            addressline={Rua Mendeleyev, 200}, 
            city={Campinas},
            postcode={13083-860}, 
            state={SP},
            country={Brazil}}


\begin{abstract}
Gravity-driven flows of liquid films are frequent in nature and industry, such as in landslides, lava flow, cooling of nuclear reactors, and coating processes. In many of these cases, the liquid is non-Newtonian and has particular characteristics. In this paper, we analyze numerically the temporal stability of films of non-Newtonian liquids falling by gravity, on the onset of instability. The liquid flows over an incline, where surface waves appear under certain conditions, and we do not fix \textit{a priori} its rheological behavior. For that, we made used of the Carreau-Yasuda model without assigning specific values to its constants, and we compute general stability solutions. The numerical strategy is based on expansions of Chebyshev polynomials for discretizing the Orr-Sommerfeld equation and boundary conditions, and a Galerkin method for solving the generalized eigenvalue problem. In addition, an Inverse Iteration method was implemented to increase accuracy and improve computational time. The result is a robust and light numerical tool capable of finding the critical conditions for different types of fluids, which we use to analyze some key fluids. We show that the outputs of the general code match previous solutions obtained for specific computations. Besides increasing our knowledge on surface-wave instabilities in non-Newtonian liquids, our findings provide a new tool for obtaining comprehensive solutions on the onset of instability.
\end{abstract}




\end{frontmatter}


\section{\label{sec0}Introduction}

Liquid films flowing under the action of gravity are common in nature and industry, happening, for example, in lava and mud flows, in the cooling of nuclear rectors, in coating processes, and when water runs down the windshield of a car. Depending on the flow conditions, surface waves known as Kapitza waves can appear and propagate downstream, their dynamics being well understood in the case of Newtonian liquids \cite{kapitsa1948wave, kapitza1949wave}. In many instances, however, the liquid is non-Newtonian and its rheology may depend on the shear rate, exhibit some plasticity, or have memory effect. In these cases, the complexity of Kapitza waves is increased by the intricate behavior of non-Newtonian fluids.

The growth of surface waves on films of Newtonian liquids has been exhaustively studied for almost a century, with experimental  \cite{kapitsa1948wave, kapitza1949wave}, analytical \cite{benney1966long, smith1990mechanism} and numerical \cite{floryan1987instabilities, Chimetta} studies, and the results converged for the well known dispersion relation (and thus wavelengths and celerities on the onset of instability) of Kapitza waves. The case is not the same for non-Newtonian liquids: given the different rheological behaviors, a general analysis of surface waves is complex and very few works inquired into it, most of works investigating analytically the instabilities appearing in specific types of liquids. For that, these works carried out linear stability analyses (LSA) using the power law \cite{ostwald1929}, Bingham \cite{bingham1916},  Carreau \cite{cross1965}, or Carreau-Yasuda \cite{yasuda1981} models for the fluid, where the constants were fixed from the beginning of the analysis (in order to model a specific fluid).

For example, Weinstein \cite{weinstein1990} investigated a multilayered flow of shear-thinning liquids down an incline by carrying out an analytical LSA with the Carreau \cite{cross1965} model. He found that in those systems the surface (Kapitza) waves behave as in Newtonian liquids, with an equivalent (effective) layer-averaged viscosity, while the interfacial waves are highly affected by the local viscosities. For the latter, a layer-averaged viscosity is not valid, the propagation of interfacial waves being thus more complex. He also showed that the growth rate of interfacial waves in shear-thinning liquids can be larger or smaller than in Newtonian liquids, and that asymptotic solutions for the velocity profile are only valid for fluids with weak shear-thinning behavior. Ng and Mei \cite{ng1994roll} and Hwang et al. \cite{hwang1994linear} investigated the surface waves appearing on a liquid film of a power-law fluid \cite{ostwald1929} flowing over an incline. The LSA of Ng and Mei \cite{ng1994roll}, based on Karman’s
momentum integral method, showed no preferential wavenumber for instabilities; however, they showed that nonlinear waves (roll waves) can exist only above a given threshold that corresponds to a minimum discharge. Interestingly, they showed that roll waves of long wavelength are suppressed for slightly non-Newtonian fluids, but they persist for highly non-Newtonian fluids. Similarly, Hwang et al. \cite{hwang1994linear} performed LSA using Karman’s
momentum integral method, and fixed different values of power-law exponent $n$. They showed that instability is enhanced by increasing the Reynolds number and decreasing the Weber number, and also by decreasing values of $n$ (in the latter case, it is accompanied by higher celerities).

Rousset et al. \cite{rousset2007} studied analytically (using a long-wave approximation) and numerically the initial instabilities of a shear-thinning fluid flowing over an incline. For that, they performed LSA by considering a Carreau fluid with fixed constants. Among other results, they showed that the critical Reynolds number is smaller for shear-thinning than for Newtonian fluids,  with a larger phase velocities, but remains proportional to the slope angle. They also found that the threshold for instability decreases with increasing the shear-thinning effects. Later, Millet et al. \cite{millet2013} investigated the stability of flows of shear-thinning two-layer liquids over an incline. For that, they solved numerically a LSA similar to that of Rousset et al. \cite{rousset2007} (Carreau model), and considered cases where the upper layer is either less or more viscous than the lower layer, both with the denser fluid on the bottom. They found that three types of instabilities can grow, leading to long-waves on the surface and long- and short-waves in the interface, and that the rheology of the lower layer greatly affects stability. In particular, they showed that the base flow and stability are only weakly affected by variations in the shear-thinning properties of the upper layer when this layer is more viscous than the lower layer. Mogilevskiy \cite{mogilevskiy2020stability} considered the effects of external excitation on shear-thinning and shear-thickening fluids flowing over an incline. He carried out a LSA for a Carreau fluid submitted to finite-amplitude perturbations (forced periodically on the incline), and obtained analytical solutions in the long-wave approximation. He showed that the forced oscillations affect the stability of a falling film, the oscillations either stabilizing or destabilizing the film flow depending on their frequency. In particular, low-frequency oscillations destabilize and stabilize the flows shear-thinning and shear-thickening fluids, respectively, while high frequency oscillations stabilize and destabilize flows of shear-thinning and shear-thickening fluids, respectively.

Different from previous works, we \cite{chimetta2020analytical} investigated analytically the base flow and the stability of a liquid
film flowing over an incline plane without fixing \textit{a priori} the exact fluid rheology. For that, we made use of long-wave approximations and considered the Carreau-Yasuda model \cite{yasuda1981}, which is a more general model and encompasses, for instance, the power-law and Carreau models. The solutions thus obtained are comprehensive, allowing for analyzing continuously the stability for different types of fluids. Chimetta and Franklin \cite{chimetta2020analytical} compared the comprehensive solution with particular solutions found in the literature, and found and excellent agreement.

Even though the analytical solutions proposed by Chimetta and Franklin \cite{chimetta2020analytical} are valid for a large range of Newtonian and non-Newtonian fluids, they are limited, in principle, to long-wave instabilities. In this paper, we solve numerically the system studied by Chimetta and Franklin \cite{chimetta2020analytical}, but without the constrain of long waves, and propose a numerical strategy for the computations. For that, we carry out temporal LSA of films of general non-Newtonian liquids falling by gravity over an incline. We use the Carreau-Yasuda model without assigning specific values to its constants, and we compute general stability solutions. The numerical strategy is based on expansions of Chebyshev polynomials for discretizing the Orr-Sommerfeld equation and boundary conditions, and a Galerkin method for solving the generalized eigenvalue problem. In addition, an Inverse Iteration method was implemented to increase accuracy and improve computational time. The result is a robust and light numerical tool capable of finding the critical conditions for different types of fluids, which we use afterwards to analyze some key fluids. We show that the outputs of the general code match previous solutions obtained for specific computations. Besides increasing our knowledge on surface-wave instabilities in non-Newtonian liquids, our findings provide a new tool for obtaining comprehensive solutions on the onset of instability. 

In the following, Sec. \ref{sec_equations} presents the model equations, Sec. \ref{sec_numerical} the numerical strategy, and Sec. \ref{sec_results} the results for Newtonian fluids, shear-thinning  and shear-thickening fluids. Finally, Sec. \ref{sec_conclusions} presents the conclusions.

\section{\label{sec_equations}Mathematical formulation}

\begin{figure}[ht]	
	\centering
	\includegraphics[width=0.7\columnwidth]{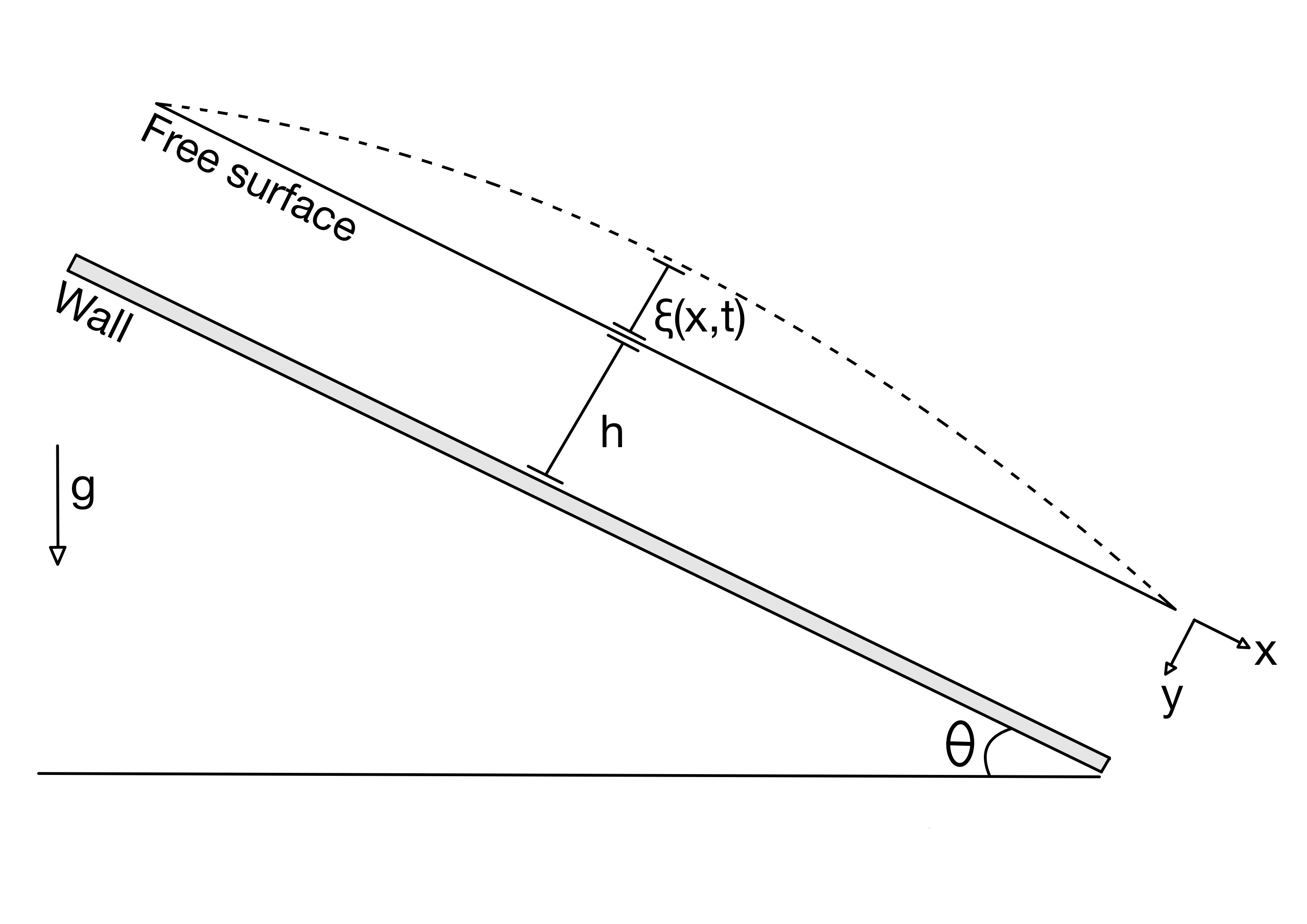}
	\caption{Layout of the falling film.}
	\label{fig:layout}
\end{figure}

We consider a liquid film of thickness $h$ driven by gravity over a plane inclined with an angle $\theta$ with respect to the horizontal. The free surface is initially flat (corresponding to the base state), with $h$ a priori unknown, the interface between liquid and gas has a surface tension $\gamma$, and the gas pressure is uniform and equal to $P_{0}$. Fig. \ref{fig:layout} presents a layout of the considered problem.

In this paper, we do not restrict the formulation to a specific type of rheology. We thus consider a generalized Newtonian fluid \cite{morrison2001understanding,macosko1994rheology}, for which the viscosity $\eta$ is a function of the shear rate $\dot{\gamma}$, 

\begin{equation}
	\underline{\underline{\tau}} = \eta (\dot{\gamma})\underline{\underline{\dot{\gamma}}} \,\,,
	\label{eq1}
\end{equation}

\noindent where $\eta(\dot{\gamma})$ is a scalar function and $\dot{\gamma} = |\underline{\underline{\dot{\gamma}}}|$. For the viscosity, we make use of the Carreau-Yasuda model \cite{yasuda1981}, which is more general than the other models (encompassing, for instance, the power-law and Carreau models), 

\begin{equation}
	\eta (\dot{\gamma}) = \eta _{\infty} + (\eta _{0} - \eta _{\infty}) [1 + (\dot{\gamma} \lambda) ^{a}] ^{\frac{n - 1}{a}} \,\, ,
	\label{eq_visc}
\end{equation}

\noindent and which has five adjusting constants: (i) $a$ controls the shape of the transition region between the zero-shear-rate plateau and the power-law region; (ii) $\lambda$ determines the values of $\dot{\gamma}$ at transitions from the zero-shear-rate plateau to power-law region and from the power-law region to that where $\eta = \eta _{\infty}$; (iii) the exponent $n$ governs the power-law region; (iv) $\eta _{\infty}$ sets the limit for large values of $\dot{\gamma}$; and (v) $\eta _0$ sets the limit for small values of $\dot{\gamma}$. Therefore, $\eta$ $\rightarrow$ $\eta _{\infty}$ as $\dot{\gamma}$ becomes large and $\eta$ $\rightarrow$ $\eta _{0}$ as $\dot{\gamma}$ becomes small.

As proposed by Weinstein \cite{weinstein1990}, we consider the length scale $h_{s}$ as the characteristic length (in the absence of an initially known value of $h$),

\begin{equation}\label{eq2}
	h_{s} = \bigg[\frac{\eta_{0}Q}{\rho g \sin(\theta)} \bigg]^{\frac{1}{3}} \,\,,
\end{equation}

\noindent where $Q$ is the volumetric flow rate by unit of width, $\rho$ is the fluid density, and $g$ is the modulus the gravity acceleration. Those quantities are then used to normalize the longitudinal $x$ and transverse $y$ coordinates, the longitudinal $u$ and transverse $v$ velocity components, the time $t$ and the pressure $p$,

\begin{equation}
	(\overline{x},\overline{y},\overline{u},\overline{v},\overline{t},\overline{p}) = \bigg(\frac{x}{h_{s}},\frac{y}{h_{s}}, \frac{uh_{s}}{Q},\frac{vh_{s}}{Q}, \frac{t Q}{h_{s}^{2}},\frac{ph_{s}^{2}}{\rho Q^{2}} \bigg) \,\,.
	\label{eq3}
\end{equation}

The dimensionless viscosity and shear rate are thus,

\begin{equation}
	\overline{\eta} (\overline{\dot{\gamma}}) = I + (1 - I) [1 + (L \dot{\gamma}) ^{a}] ^{\frac{n - 1}{a}} \,\,,
	\label{eq4}
\end{equation}

\begin{equation}
	\overline{\dot{\gamma}} = \bigg\{ 2 \bigg(\frac{\partial \overline{u}}{\partial \overline{x}} \bigg)^{2} + \bigg[\bigg(\frac{\partial \overline{u}}{\partial \overline{y}} \bigg)^{2} + 2\frac{\partial \overline{u}}{\partial \overline{y}} \frac{\partial \overline{v}}{\partial \overline{x}} + \bigg(\frac{\partial \overline{v}}{\partial \overline{x}} \bigg)^{2} \bigg] + 2 \bigg(\frac{\partial \overline{v}}{\partial \overline{y}} \bigg)^{2} \bigg\}^{\frac{1}{2}} \,\,,
	\label{eq5}
\end{equation}

\noindent where $I = \eta_{\infty} / \eta_{0}$ is the ratio between the limits of viscosity (large over small) and $L = \lambda Q / h_{s} ^{2}$ is a relaxation time. The dimensionless viscosity is $\overline{\eta} = \eta / \eta_{0}$.

\subsection{\label{sec1sub1}Base state}

For a two-dimensional (2D) flow, the dimensionless conservation of mass and momentum are described by Eqs. $\ref{eq6}$, $\ref{eq7}$ and $\ref{eq8}$,

\begin{equation}
	\frac{\partial \overline{u}}{\partial \overline{x}} + \frac{\partial \overline{v}}{\partial \overline{y}} = 0 \,\,,
	\label{eq6}
\end{equation}

\begin{equation}
	\frac{\partial \overline{u}}{\partial \overline{t}} + \overline{u} \frac{\partial \overline{u}}{\partial \overline{x}} + \overline{v} \frac{\partial \overline{u}}{\partial \overline{y}} = - \frac{\partial \overline{p}}{\partial \overline{x}} + \frac{1}{Re} \bigg(\frac{\partial \overline{\tau} _{xx}}{\partial \overline{x}} + \frac{\partial \overline{\tau} _{xy}}{\partial \overline{y}}\bigg) + \frac{1}{Fr _{x} ^{2}} \,\,,
	\label{eq7}
\end{equation}

\begin{equation}
	\frac{\partial \overline{v}}{\partial \overline{t}} + \overline{u} \frac{\partial \overline{v}}{\partial \overline{x}} + \overline{v} \frac{\partial \overline{v}}{\partial \overline{y}} = - \frac{\partial \overline{p}}{\partial \overline{y}} + \frac{1}{Re} \bigg(\frac{\partial \overline{\tau} _{xy}}{\partial \overline{x}} + \frac{\partial \overline{\tau} _{yy}}{\partial \overline{y}}\bigg) + \frac{1}{Fr _{y} ^{2}} \,\,,
	\label{eq8}
\end{equation}

\noindent where the dimensionless stress tensor components are given by Eqs. $\ref{eq9}$, $\ref{eq10}$ and $\ref{eq11}$,

\begin{equation}
	\overline{\tau} _{xx} = 2 \overline{\eta} (\overline{\dot{\gamma}}) \frac{\partial \overline{u}}{\partial \overline{x}} \,\,,
	\label{eq9}
\end{equation}

\begin{equation}
	\overline{\tau} _{yy} = 2 \overline{\eta} (\overline{\dot{\gamma}}) \frac{\partial \overline{v}}{\partial \overline{y}} \,\,,
	\label{eq10}
\end{equation}

\begin{equation}
	\overline{\tau} _{xy} = \overline{\eta} (\overline{\dot{\gamma}}) \bigg[\frac{\partial \overline{u}}{\partial \overline{y}} + \frac{\partial \overline{v}}{\partial \overline{x}}\bigg] \,\,,
	\label{eq11}
\end{equation}

\noindent and the Reynolds and Froude numbers in Eqs. $\ref{eq6}$ to $\ref{eq8}$ are given by

\begin{equation}
	(Fr_{x}, Fr_{y}, Re) = \bigg( \sqrt{\frac{Q ^{2}}{gh_{s} ^{3}\sin(\theta)}}, \sqrt{\frac{Q ^{2}}{gh_{s} ^{3}\cos(\theta)}}, \frac{\rho Q}{\eta _{0}} \bigg) \,\,.
	\label{eq12}
\end{equation}

For the base state, the flow is parallel and steady, with a velocity profile equals to $\overline{U}$. Velocity in the normal direction is equal to zero and the base flow is a function of $\overline{y}$ only. For the pressure gradient, only the normal component is different from zero. The boundary conditions are zero shear at the free surface and no-slip at the wall, both corresponding to $\overline{y} = 0$ and $\overline{y} = \overline{h}$ respectively. Under these assumptions, it is possible to write,

\begin{equation}
	\overline{U}(\overline{h}) = 0 \,\,,
	\label{eq13}
\end{equation}

\color{black}
\begin{equation}
	\bigg\{ I + (1 - I) \bigg[1 + \bigg(L \left| \frac{d \overline{U}}{d \overline{y}} \right| \bigg) ^{a} \bigg] ^{\frac{n - 1}{a}} \bigg \} \frac{d \overline{U}}{d \overline{y}} = - \overline{y} \   \ for \   \ \overline{y} \in [0;\overline{h}] \,\,,
	\label{eq14}
\end{equation}
\color{black}

\noindent where $\overline{h} = h/h _{s}$. Equation \ref{eq14} is obtained by considering the dimensionless flow rate equal to unity,

\begin{equation}
	\int_{0}^{\overline{h}} \overline{U}d \overline{y} = 1 \,\,.
	\label{eq15}
\end{equation}

Equations \ref{eq13}, \ref{eq14} and \ref{eq15} establish a nonlinear problem for the film thickness and the velocity profile, with no general analytical solution. However, an approximate solution was obtained through an asymptotic analysis by Chimetta and Franklin \cite{chimetta2020analytical}.

\subsection{\label{sec1sub2}Perturbations}

For parallel flows of Newtonian fluids, the Squire's theorem \cite{squire1933} states that the most unstable modes are 2D, a result that was proved valid for liquids with free-surface \cite{yih1955} and in stratified flows \cite{hesla1986}. Later, Nouar et al. \cite{nouar2007} showed that two-dimensional instabilities emerge in parallel flows of non-Newtonian fluids (modelled by the Carreau model), indicating that the Squire's Theorem would be also valid for non-Newtonian liquids. Therefore, as in  Chimetta and Franklin \cite{chimetta2020analytical}, we assume that the Squire's Theorem is valid and two-dimensional perturbations are dominant.

We consider small perturbations for the longitudinal $\hat{u}$ and transverse $\hat{v}$ components of velocity, for the pressure $\hat{p}$, and and for the interface $\hat{\xi}$. The velocity and pressure fields become then $\overline{u} = \overline{U} + \hat{u}$, $\overline{v} = 0 + \hat{v}$, and $\overline{p} = \overline{P} + \hat{p}$, and the interface position $\overline{\xi} = 0 + \hat{\xi}$. Also, all products between perturbations must be neglected in a linear stability analysis. In two dimensions, it is possible to use stream functions for the velocity field,

\begin{equation}\label{eq16}
	(\hat{u},\hat{v}) = \bigg(\frac{\partial \hat{\Psi}}{\partial \overline{y}}, - \frac{\partial \hat{\Psi}}{\partial \overline{x}} \bigg) \,\,,
\end{equation}

\noindent where $\hat{\Psi}$ is the perturbation of the streamline function ($\Psi = \overline{\Psi} + \hat{\Psi}$). By inserting the perturbations in Eqs. \ref{eq6} to \ref{eq8} and linearizing, the expected solutions for the perturbations are are plane waves, given by Eqs. \ref{eq17} and \ref{eq18},

\begin{equation}
	\hat{\Psi}(\overline{x},\overline{y},\overline{t}) = \tilde{\Psi}(\overline{y}) e ^{i \alpha (\overline{x} - c \overline{t})}
	\label{eq17}
\end{equation}

\begin{equation}
	\hat{\xi}(\overline{x},\overline{t}) = \tilde{\xi} e ^{i \alpha (\overline{x} - c \overline{t})}
	\label{eq18}
\end{equation}

\noindent where $\alpha = kh_{s} \in \mathbb{R}$, $k$ being the wave number, and $c = \omega h_{s}k^{-1}Q^{-1} \in \mathbb{C}$, $\omega$ corresponding to the complex frequency (which defines the temporal stability approach). We consider $c = c_{r} + i c_{i}$, where $\sigma = \alpha c_{i}$ corresponds to the growth rate and $c_{r}$ is the phase velocity. The system is stable when $c_{i} < 0$ and linearly unstable if $c_{i} > 0$, $c_{i} = 0$ representing neutral stability. By inserting Eqs. \ref{eq16}, \ref{eq17} and \ref{eq18} into Eqs. \ref{eq6}, \ref{eq7} and \ref{eq8}, we obtain the equivalent of the Orr-Sommerfeld equation for a Carreau-Yasuda fluid,

\color{black}
\begin{equation*}
	(D ^{2} + \alpha ^{2})[D ^{2} \overline{\epsilon}_{t} + 2D\overline{\epsilon}_{t}D + \overline{\epsilon}_{t}(D ^{2} + \alpha ^{2})]\tilde{\Psi} - 4\alpha ^{2}D(\overline{\eta}D\tilde{\Psi}) =
\end{equation*}
\color{black}
\begin{equation}
	= i\alpha Re [(\overline{U}-c)(D ^{2} - \alpha ^{2}) - D ^{2}\overline{U}]\tilde{\Psi} \,\,,
	\label{eq19}
\end{equation}

\noindent where $D ^{j} = \frac{\partial ^{j}}{\partial \overline{y} ^{j}}$. \textcolor{black}{The term $\overline{\epsilon}_{t}$ is obtained} as described in \ref{tangent_viscosity} and can be written as,

\color{black}
\begin{equation}
	\overline{\epsilon}_{t} = I + (1 - I)\bigg[1 + n \bigg(L \left| \frac{\partial \overline{U}}{\partial \overline{y}} \right| \bigg)^{a} \bigg] \bigg[1 + \bigg(L \left| \frac{\partial \overline{U}}{\partial \overline{y}} \right| \bigg)^{a} \bigg]^{\frac{n-a-1}{a}} \,\,.
	\label{eq20}
\end{equation}
\color{black}

With that, the no-slip conditions at the wall for the longitudinal and transverse components of velocity are given, respectively, by Eqs.\ref{eq21} and \ref{eq22},

\begin{equation}
	\tilde{\Psi}(\overline{h}) = 0 \,\,,
	\label{eq21}
\end{equation}

\begin{equation}
	D\tilde{\Psi}(\overline{h}) = 0 \,\,,
	\label{eq22}
\end{equation}

The boundary conditions at the free surface ($y = 0$) are the kinematic condition, which represents the impermeability of the interface, and the continuity of the tangential and normal stresses through the interface, which are related to the viscous effect and the Laplace-Young equation. These conditions are given by Eqs. \ref{bc1}-\ref{bc3}, respectively,

\begin{equation}\label{bc1}
	\frac{\partial \overline{\xi}}{\partial \overline{t}} + \overline{u}\frac{\partial \overline{\xi}}{\partial \overline{x}} - \overline{v} = 0 \,\,,
\end{equation}

\begin{equation}\label{bc2}
	\overline{\tau}_{xy} - \overline{\xi} = 0 \,\,,
\end{equation}

\begin{equation}\label{bc3}
	-2\overline{\tau}_{xy}\frac{\partial \overline{\xi}(\overline{x},\overline{t})}{\partial \overline{x}} + \overline{\tau}_{yy} - \overline{p}Re + \frac{1}{We}\frac{\partial^{2} \overline{\xi}}{\partial \overline{x}^{2}} = 0 \,\,.
\end{equation}

\noindent Inserting Eqs. \ref{eq16}-\ref{eq18} into Eqs. \ref{bc1}-\ref{bc3} results in

\begin{equation}\label{bc4}
	\tilde{\Psi} - (c - \overline{U})\tilde{\xi} = 0 \,\,,
\end{equation}

\begin{equation}\label{bc5}
	\textcolor{black}{\overline{\epsilon}_{t}(D^{2} + \alpha ^{2})\tilde{\Psi} - \tilde{\xi} = 0 \,\,,}
\end{equation}

\begin{equation*}
	\textcolor{black}{i\alpha Re \Biggl[(c-\overline{U})\frac{\partial \tilde{\Psi}}{\partial \overline{y}} + \frac{\partial \overline{U}}{\partial \overline{y}}\tilde{\Psi}\Biggl] - 4\alpha ^{2}\overline{\eta}\frac{\partial \tilde{\Psi}}{\partial \overline{y}} + \frac{\partial \overline{\epsilon}_{t}}{\partial \overline{y}}\Biggl(\frac{\partial ^{2}\tilde{\Psi}}{\partial \overline{y}^{2}} + \alpha ^{2}\tilde{\Psi}\Biggl) }
\end{equation*}
\begin{equation}\label{bc6}
	\textcolor{black}{+ \overline{\epsilon}_{t}\Biggl(\frac{\partial ^{3}\tilde{\Psi}}{\partial \overline{y}^{3}}+\alpha ^{2}\frac{\partial \tilde{\Psi}}{\partial \overline{y}}\Biggl) + i\alpha \Biggl( \cot \theta + \frac{\alpha ^{2}}{We}\Biggl)\tilde{\xi} = 0 \,\,.}
\end{equation}

By inserting Eq. \ref{bc5} into Eqs. \ref{bc4} and \ref{bc6}, we obtain the boundary conditions at the free-surface, given by Eqs. \ref{eq23} (kinematic) and \ref{eq24} (dynamic),

\begin{equation}\label{eq23}
	[1 + (\overline{U} - c)(D^{2}+\alpha ^{2})]\tilde{\Psi}(0) = 0 \,\,, 
\end{equation}

\begin{eqnarray*}
	i\alpha Re[(c - \overline{U})D + D\overline{U}]\tilde{\Psi}(0) - 4\alpha ^{2}\overline{\eta}D \tilde{\Psi}(0) + (D^{2}+\alpha ^{2}) \,\,,
\end{eqnarray*}
\begin{eqnarray}\label{eq24}
	\textcolor{black}{\bigg[D\overline{\epsilon}_{t} + \overline{\epsilon}_{t}D + i\alpha \overline{\epsilon}_{t} \bigg(\cot \theta + \frac{\alpha ^{2}}{We_{m}} \bigg) \bigg]\tilde{\Psi}(0) = 0 \,\,,}
\end{eqnarray}

\noindent where $We _{m} = \eta _{0} Q (h_{s} \gamma )^{-1}$ is a modified Weber number. Equations \ref{eq19}, \ref{eq21}, \ref{eq22}, \ref{eq23} and \ref{eq24} establish a generalized eigenvalue problem for the complex frequency $c$.

\section{\label{sec_numerical}Numerical strategy}

We implemented a spectral method \cite{fletcher1984computational, boyd1989chebyshev} to solve Eqs. \ref{eq20} to \ref{eq24}, by making use of weighted residuals methods, which we describe briefly in \ref{appendix_weighted_method}.

\subsection{\label{sec2sub1} Numerical formulation for the base state}

In order to use Chebyshev polynomials, we introduce the transformation

\begin{equation}
	\overline{z} = \frac{2\overline{y}}{\overline{h}} - 1 \,\,,
	\label{eq25}
\end{equation}

\noindent which transfers the domain $\overline{y} \in [0; \overline{h}]$ to $\overline{z} \in [-1; 1]$. With that, Eqs. \ref{eq13}--\ref{eq15} become

\begin{equation}\label{eq26}
	\overline{U}(1) = 0 \,\,,
\end{equation}

\begin{equation*}
	\bigg\{ I + (1 - I) \Biggl[1 + \Biggl(\frac{2L}{\overline{h}} \frac{d \overline{U}}{d \overline{z}} \Biggl) ^{a} \Biggl] ^{\frac{n - 1}{a}} \bigg \}\frac{2}{\overline{h}} \frac{d \overline{U}}{d \overline{z}}
\end{equation*}
\begin{equation}\label{eq27}
	=\frac{-\overline{h}(\overline{z} + 1)}{2} \ \text{   for   } \ \overline{z} \in [-1;1] \,\,,
\end{equation}

\begin{equation}\label{eq28}
	\int_{-1}^{1}  \overline{U}\frac{\overline{h}}{2}d \overline{z} = 1 \,\,.
\end{equation}

We solve this system in MatLab with the built-in function bvp4c, which is a finite-difference discretization of a three-stage Lobatto formula \cite{kierzenka2001bvp, shampine2000solving}. Since Eqs. \ref{eq26}--\ref{eq28} correspond to a non-linear problem with an undetermined parameter $\overline{h}$ and an integral boundary condition, it is necessary to write three distinct functions within the code: one that represents a system of first-order equations, one for the boundary conditions, and one for the initial guess. By considering $\frac{d\overline{g}}{d\overline{z}} = \overline{U}$; $\frac{d^{2}\overline{g}}{d\overline{z}^{2}} = \frac{d\overline{U}}{d\overline{z}}$, rewriting Eq. \ref{eq28} as $\overline{g}(1) - \overline{g}(- 1) = \frac{2}{\overline{h}}$, and using the notation $y_{1} = \overline{g}$ and $y_{2} = \frac{d \overline{g}}{d \overline{z}}$, we obtain Eqs. \ref{eq34}--\ref{eq37}

\begin{equation}\label{eq34}
	y_{2}(1) = 0 \,\,,
\end{equation}

\begin{equation}\label{eq35}
	\frac{dy_{2}}{d\overline{z}} = \frac{\frac{-\overline{h}^{2}(\overline{z} + 1)}{4}}{\bigg\{ I + (1 - I) \Biggl[1 + \Biggl(\frac{2L}{\overline{h}} \frac{dy_{2}}{d\overline{z}} \Biggl) ^{a} \Biggl] ^{\frac{n - 1}{a}} \bigg \} } \,\,,
\end{equation}

\begin{equation}\label{eq36}
	y_{1}(1) - y_{1}(-1) = \frac{2}{\overline{h}} \,\,,
\end{equation}

\begin{equation}\label{eq37}
	y_{1}(-1) = 0 \,\,.
\end{equation}

Equations \ref{eq34}--\ref{eq37} are solved to obtain the velocity profile $\overline{U}$ and the liquid film thickness $\overline{h}$, in addition to obtaining $\overline{\eta}$, \textcolor{black}{$\overline{\epsilon}_{t}$} and all the derivatives. For that, their Chebyshev coefficients are computed with the open-source package Chebfun \cite{driscoll2014chebfun, trefethen2019approximation}, which is used together with the bvp4c function.

\subsection{\label{sec2sub2} Numerical formulation for the perturbations}

\indent Inserting the transformation given by Eq. \ref{eq25} into Eq. \ref{eq19} leads to

\begin{equation*}
	\textcolor{black}{\Biggl(\frac{4}{\overline{h}^{2}}D^{2}+\alpha ^{2}\Biggl)\Biggl[\frac{4}{\overline{h}^{2}}D^{2}\overline{\epsilon}_{t} + \frac{8}{\overline{h}^{2}}D\overline{\epsilon}_{t}D + \overline{\epsilon}_{t}\Biggl(\frac{4}{\overline{h}^{2}}D^{2}+\alpha ^{2}\Biggl)\Biggl]\tilde{\Psi} }
\end{equation*}
\begin{equation}\label{eq38}
	- \frac{16}{\overline{h}^{2}}\alpha ^{2}D(\overline{\eta}D\tilde{\Psi}) = i\alpha Re \Biggl[(\overline{U} - c)\Biggl(\frac{4}{\overline{h}^{2}}D^{2} - \alpha ^{2}\Biggl) - \frac{4}{\overline{h}^{2}}D^{2}\overline{U}\Biggl]\tilde{\Psi} \,\,,
\end{equation}

\noindent which is the Orr-Sommerfeld equation within $\overline{z} \in [-1; 1]$. Then, inserting Eq. \ref{eq25} into Eqs. \ref{eq4} and \ref{eq20} gives

\begin{equation}
	\overline{\eta}(\overline{z}) = I + (1 - I)\Biggl[1 + \Biggl(\frac{2L}{\overline{h}} D \overline{U} \Biggl)^{a}\Biggl]^{\frac{n-1}{a}} \,\,,
	\label{eq39}
\end{equation}

\begin{equation}
	\textcolor{black}{\overline{\epsilon}_{t}(\overline{z}) = I + (1 - I)\bigg[1 + n \bigg(\frac{2L}{\overline{h}} D \overline{U} \bigg)^{a} \bigg] \bigg[1 + \bigg(\frac{2L}{\overline{h}} D \overline{U} \bigg)^{a} \bigg]^{\frac{n-1}{a}-1} \,\,.}
	\label{eq40}
\end{equation}

By applying the same procedure for the no-slip conditions at the wall (Eqs. \ref{eq21} and \ref{eq22}), we obtain

\begin{equation}
	\tilde{\Psi}(\overline{z}) = 0 \,\,,
	\label{eq41}
\end{equation}

\begin{equation}
	D\tilde{\Psi}(\overline{z}) = 0 \,\,,
	\label{eq42}
\end{equation}

\noindent at $\overline{z} = 1$. For the boundary conditions at the free-surface (Eqs. \ref{eq23} and \ref{eq24}), we find

\begin{equation}\label{eq43}
	\textcolor{black}{\Biggl[1 + \overline{\epsilon}_{t}(\overline{U} - c)\Biggl(\frac{4}{\overline{h}^{2}}D^{2} + \alpha ^{2}\Biggl)\Biggl]\tilde{\Psi} = 0 \,\,,}
\end{equation}

\begin{equation*}
	i\alpha Re\Biggl[\frac{2}{\overline{h}}(c-\overline{U})D + \frac{2}{\overline{h}}D\overline{U}\Biggl]\tilde{\Psi} - \frac{8}{\overline{h}}\alpha ^{2}\overline{\eta}D\tilde{\Psi}
\end{equation*}
\begin{equation}\label{eq44}
	\textcolor{black}{+ \Biggl(\frac{4}{\overline{h}^{2}}D^{2} + \alpha ^{2}\Biggl)\Biggl[\frac{2}{\overline{h}}\overline{\epsilon}_{t}D + \frac{2}{\overline{h}}D\overline{\epsilon}_{t} + i\alpha \overline{\epsilon}_{t}\Biggl(\cot \theta + \frac{\alpha ^{2}}{We}\Biggl)\Biggl]\tilde{\Psi} = 0 \,\,,}
\end{equation}

\noindent at $\overline{z} = -1$. To solve the Eqs. \ref{eq38} to \ref{eq44}, we expand the perturbed streamfunction as a sum of products between Chebyshev coefficients $\Phi _{k}$ and polynomials $T_{k}$,

\begin{equation}
	\tilde{\Psi}(\overline{z}) = \sum_{k = 0}^{N} \Phi_{k} T_{k}(\overline{z}) \,\,,
	\label{eq45}
\end{equation}

\noindent where $T_k (\overline{z}) = \cos (k\ \arccos \overline{z})$ and $k \in \{\mathbb{Z} | k \geqslant 0 \}$. Proceeding as described in \ref{appendix_weighted_method}, we insert Eq. \ref{eq45} into Eq. \ref{eq38}, with a product using the Chebyshev polynomial $T_{j}(\overline{z})$ and a weight function $\hat{w} = (1 - \overline{z}^{2})^{-\frac{1}{2}}$. This procedure leads to

\begin{equation*}
	\textcolor{black}{\sum_{j = 0}^{N} \sum_{k = 0}^{N} \Biggl[ \frac{16}{\overline{h}^{4}}\left< T _{j}(\overline{z}),\overline{\epsilon}_{t}D^{4}T _{k}(\overline{z})\right> + \frac{8}{\overline{h}^{2}}\alpha ^{2} \left<T _{j}(\overline{z}),\overline{\epsilon}_{t}D^{2}T _{k}(\overline{z}) \right>}
\end{equation*}
\begin{equation*}
	\textcolor{black}{+ \alpha ^{4} \left<T _{j}(\overline{z}),\overline{\epsilon}_{t} T _{k}(\overline{z})\right> + \frac{32}{\overline{h}^{4}} \left<T _{j}(\overline{z}),D\overline{\epsilon}_{t}D^{3}T _{k}(\overline{z})\right> }
\end{equation*}
\begin{equation*}
	\textcolor{black}{+ \frac{8}{\overline{h}^{2}}\alpha ^{2} \left<T _{j}(\overline{z}),D\overline{\epsilon}_{t}D T _{k}(\overline{z})\right> + \frac{16}{\overline{h}^{4}} \left<T _{j}(\overline{z}),D^{2}\overline{\epsilon}_{t}D^{2}T _{k}(\overline{z})\right> }
\end{equation*}
\begin{equation*}
	\textcolor{black}{+ \frac{4}{\overline{h}^{2}}\alpha ^{2} \left<T _{j}(\overline{z}),D^{2}\overline{\epsilon}_{t}T _{k}(\overline{z})\right> - \frac{16}{\overline{h}^{2}}\alpha ^{2} \left<T _{j}(\overline{z}),D\overline{\eta}DT _{k}(\overline{z})\right>}
\end{equation*}
\begin{equation*}
	- \frac{16}{\overline{h}^{2}}\alpha ^{2} \left<T _{j}(\overline{z}),\overline{\eta}D^{2}T _{k}(\overline{z})\right> - \frac{4}{\overline{h}^{2}}i\alpha Re \left<T _{j}(\overline{z}),\overline{U}D^{2}T _{k}(\overline{z})\right>
\end{equation*}
\begin{equation*}
	+ i\alpha^{3} Re \left<T _{j}(\overline{z}),\overline{U}T _{k}(\overline{z})\right> + \frac{4}{\overline{h}^{2}}i\alpha Re \left<T _{j}(\overline{z}),D^{2}\overline{U}T _{k}(\overline{z})\right> \Biggl] \Phi _{k}
\end{equation*}
\begin{equation*}
	= c \sum_{j = 0}^{N} \sum_{k = 0}^{N} \Biggl[ -\frac{4}{\overline{h}^{2}} i \alpha Re \left<T _{j}(\overline{z}),D^{2}T _{k}(\overline{z})\right>
\end{equation*}
\begin{equation}\label{eq46}
	+ i \alpha^{3} Re  \left<T _{j}(\overline{z}),T _{k}(\overline{z})\right> \Biggl] \Phi _{k} \,\,,
\end{equation}

\noindent where $\left<f,g\right>$ is the scalar product between $f$ and $g$ (\ref{appendix_weighted_method}). For the no-slip conditions at the solid wall, we obtain

\begin{equation}
	\sum_{k=0}^{N} \Phi_{k} T_{k}(1) = 0 \,\,,
	\label{eq47}
\end{equation}

\begin{equation}
	\sum_{k=0}^{N} \Phi_{k}D T_{k}(1) = 0 \,\,,
	\label{eq48}
\end{equation}

\noindent and for the free surface,

\begin{equation*}
	\textcolor{black}{\sum_{k=0}^{N} \Biggl[ T_{k}(-1) +\frac{4}{\overline{h}^{2}}\overline{\epsilon}_{t}\overline{U}D^{2}T_{k}(-1) + \alpha ^{2}\overline{\epsilon}_{t}\overline{U} T_{k}(-1)\Biggl] \Phi_{k}}
\end{equation*}
\begin{equation}\label{eq49}
	\textcolor{black}{= c \sum_{k=0}^{N} \Biggl[ \frac{4}{\overline{h}^{2}}\overline{\epsilon}_{t}D^{2}T_{k}(-1) + \alpha ^{2}\overline{\epsilon}_{t}T_{k}(-1) \Biggl] \Phi_{k} \,\,,}
\end{equation}

\begin{equation*}
	\textcolor{black}{\sum_{k=0}^{N} \Biggl\{-\frac{8}{\overline{h}}\alpha ^{2} \overline{\eta}DT_{k}(-1) + \frac{8}{\overline{h}^{3}}D\overline{\epsilon}_{t}D ^{2}T_{k}(-1) + \frac{2}{\overline{h}}\alpha ^{2}D\overline{\epsilon}_{t}T_{k}(-1) }
\end{equation*}
\begin{equation*}
	\textcolor{black}{+ \frac{8}{\overline{h}^{3}} \overline{\epsilon}_{t}D^{3}T_{k}(-1) + \frac{2}{\overline{h}} \alpha ^{2}\overline{\epsilon}_{t}DT _{k}(-1) + i \Biggl[ - \frac{2}{\overline{h}} \alpha Re \overline{U} DT_{k}(-1)}
\end{equation*}
\begin{equation*}
	\textcolor{black}{+ \frac{2}{\overline{h}} \alpha Re D\overline{U}T_{k}(-1) + \frac{4}{\overline{h}^{2}} \alpha \overline{\epsilon}_{t} \cot \theta D^{2}T_{k}(-1) + \alpha ^{3} \cot \theta \overline{\epsilon}_{t} T_{k}(-1) }
\end{equation*}
\begin{equation*}
	\textcolor{black}{+ \frac{4 \alpha ^{3}}{\overline{h} ^{2} We}\overline{\epsilon}_{t}D ^{2}T_{k}(-1) + \frac{\alpha ^{5}}{We}\overline{\epsilon}_{t}T_{k}(-1)\Biggl] \Biggl\} \Phi_{k} }
\end{equation*}
\begin{equation}\label{eq50}
	= c \sum_{k=0}^{N} \Biggl\{ - \frac{2}{\overline{h}}i\alpha Re DT_{k}(-1)\Biggl\}\Phi_{k} \,\,,
\end{equation}

Equation \ref{eq46}, as well as the functions $\overline{U} $, $\overline{\eta}$, \textcolor{black}{$\overline{\epsilon}_{t}$ and their derivatives,} form a matrix with order $N$ x $N$ whose discretization leads to

\begin{equation}
	[\textbf{A}]_{N\times N} \vec{a} = c [\textbf{B}]_{N \times N} \vec{a} \,\,,
	\label{eq51}
\end{equation}

\noindent where $N$ represents the number of Chebyshev polynomials used in the discretization process, $A, B \in {\cal M}_{m _{x} n} ^{\mathbb{C}}$ with products of Chebychev polynomials, and $\vec{a}$ is the eigenvector (matrices, together with the numerical scripts, are available on an open repository \cite{chimettamendeley2023}). The discretized boundary conditions (Eqs. \ref{eq47}--\ref{eq50} can be written as

\begin{equation}
	[\textbf{A}]_{1 \times N} \vec{a} = c [\textbf{B}]_{1 \times N} \vec{a} \,\,,
	\label{eq52}
\end{equation}

\indent We insert Eq. \ref{eq52} as Eq. \ref{eq51} (as the last four rows of $\textbf{A}$ and $\textbf{B}$), ending with a generalized eigenvalue problem. To solve the eigenvalue problem, we use the MatLab’s built-in function $eig$, which makes use of a QZ algorithm by default.

\subsection{\label{sec2sub3} The inverse iteration method}

\indent When the eigenvalue problem is solved, a spectrum of eigenvalues and eigenvectors is generated. To check the convergence of the physical solution, a few tests are required while the number of Chebyshev polynomials increases on each test. From this process, some difficulties arise. The first is the computational cost to produce the results, especially the stability diagram. The second problem involves how MatLab processes and stores the results in the arrays. Every time that a parameter or the number of Chebyshev polynomials are adjusted, the position of the converged eigenvalue and the eigenvector spectrum change. To overcome this problem, an inverse iteration method can be used. This method boosts the precision of the eigenvalue solution while decreasing deeply the computational time, once solving the eigenvalue problem using \textit{eig} is not required in each iteration. Therefore, the first solution, obtained with the function \textit{eig}, works as an initial guess for the inverse iteration, which tracks the next physical solution, eliminating the necessity of processing the complete spectrum \cite{hossain2011convection}. Following that, a version of the inverse iteration algorithm used in this work is presented.

\begin{algorithm}
	\caption{Inverse iteration algorithm.}
	\begin{algorithmic}[1]
		\STATE $\textbf{A}$,$\textbf{B}$ : Left and right matrices from the generalized eigenvalue problem
		\STATE $\sigma_{0}$ : Initial approximation for the eigenvalue
		\STATE $\textbf{z}_{k}$ : Approximation for the eigenvector in each iteration (for $k$ = 0, $\textbf{z}_{0}$ is the initial approximation)
		\STATE $\textbf{z}_{k+1}$ : Normalization of the approximate eigenvector
		\STATE $\textbf{w}_{k+1}$ : Computed eigenvector in each iteration based on $\sigma_{0}$ and $\textbf{z}_{k}$
		\STATE $ \|\textbf{w}_{k+1} \|_{2}$ : L2-norm of the eigenvector $\textbf{w}_{k+1}$ 
		\STATE $p_{k+1}$ : Inverse of the inner product between the computed eigenvector $\textbf{w}_{k+1}$ and the approximation $\textbf{z}_{k}$ (for $k$ = 0,  $p_{0}$ = 0 as initial parameter)
		\STATE \textcolor{black}{$\kappa$ : Test parameter for reaching the desired convergence threshold (considered $10^{-10}$ in our computations)}
		\WHILE{$k = 0,1,2,3,...$}
		\STATE Solve $(\textbf{A} - \sigma_{0}\textbf{B})\textbf{w}_{k+1} = \textbf{B}\textbf{z}_{k}$
		\STATE Compute $p_{k+1} = \left<\textbf{w}_{k+1},\textbf{z}_{k}\right>^{-1}$
		\IF{\textcolor{black}{$|p_{k+1}-p_{k}| > \kappa$}}
		\STATE Compute the normalized eigenvector $\textbf{z}_{k+1} = \textbf{w}_{k+1}/ \|\textbf{w}_{k+1} \|_{2}$
		\STATE RETURN TO STEP 10
		\ELSE
		\STATE Compute the eigenvalue $\sigma = \sigma_{0} + p_{k+1}$
		\STATE Compute the normalized eigenvector $\textbf{z}_{k+1} = \textbf{w}_{k+1}/ \|\textbf{w}_{k+1} \|_{2}$
		\STATE STOP
		\ENDIF
		\ENDWHILE
	\end{algorithmic}
\end{algorithm}

\section{\label{sec_results} Results}

Our numerical computations do not suppose \textit{a priori} the specific fluid rheology, being valid for any fluid obeying the Carreau-Yasuda model. Therefore, different from previous works, we can gradually vary the type of fluid and investigate how stability changes. We inquire next into the base state and perturbations of shear-thinning, Newtonian and shear-thickening fluids. For that, we vary gradually the parameters $a$ and $n$ and plot the solutions.

\subsection{\label{sec3sub1} Base state}

\begin{figure}[!htb]
	\begin{center}
		\includegraphics[width=0.7\columnwidth]{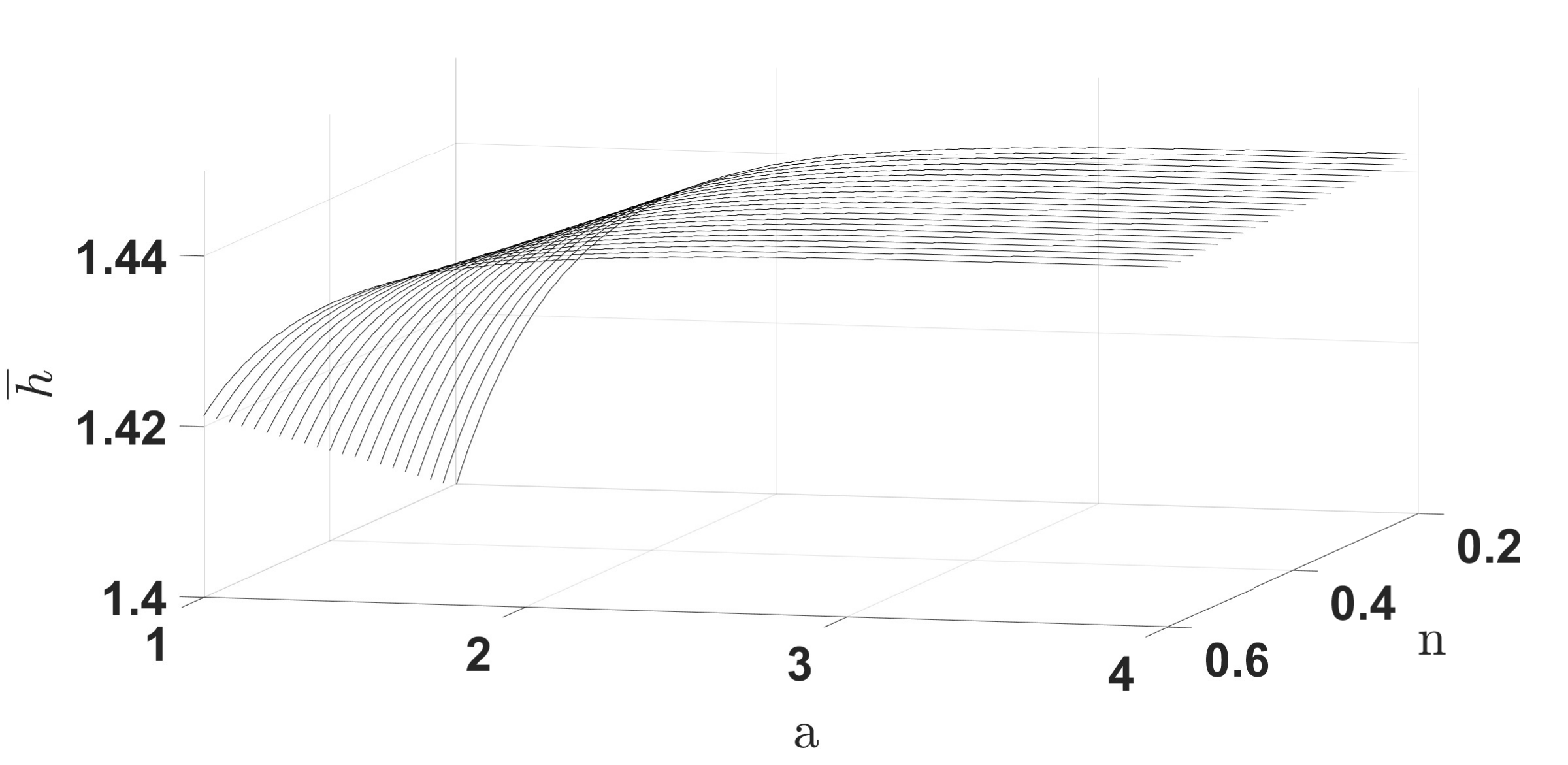}
	\end{center}
	\caption{\textcolor{black}{Numerical result of $\overline{h}$ for shear-thinning fluids with 0.2 $\leq$ $n$ $\leq$ 0.6, 1 $\leq$ $a$ $\leq$ 4, $I$ = 0 and $L$ = 0.4.}}
	\label{fig10}
\end{figure}

\begin{figure}[!htb]
	\begin{center}
		\includegraphics[width=0.7\columnwidth]{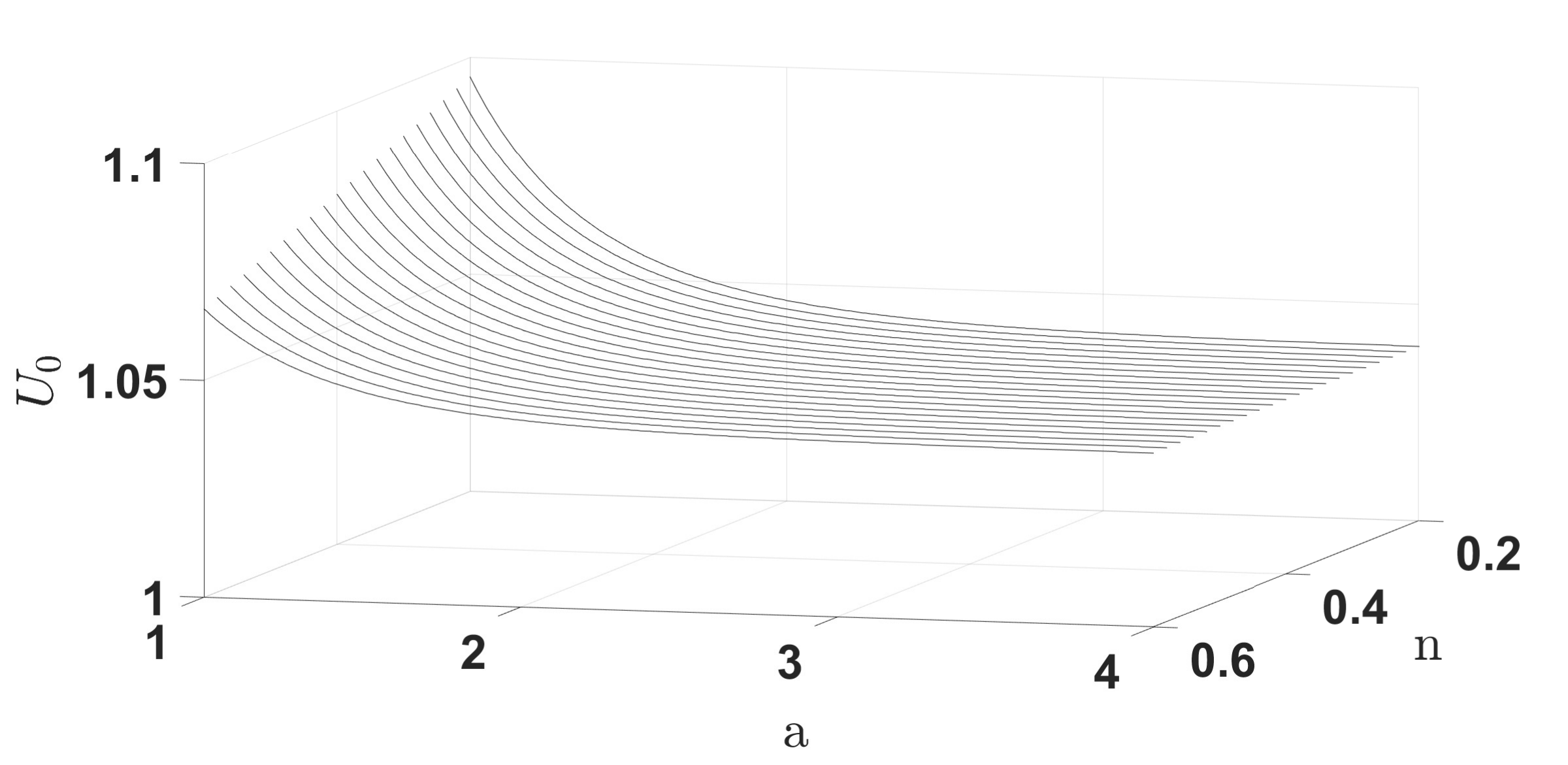}
	\end{center}
	\caption{\textcolor{black}{Numerical result of $U_{0}$ (surface velocity) for shear-thinning fluids with 0.2 $\leq$ $n$ $\leq$ 1, 1 $\leq$ $a$ $\leq$ 4, $I$ = 0 and $L$ = 0.4.}}
	\label{fig11}
\end{figure}

\begin{figure}[!htb]
	\begin{center}
		\includegraphics[width=0.7\columnwidth]{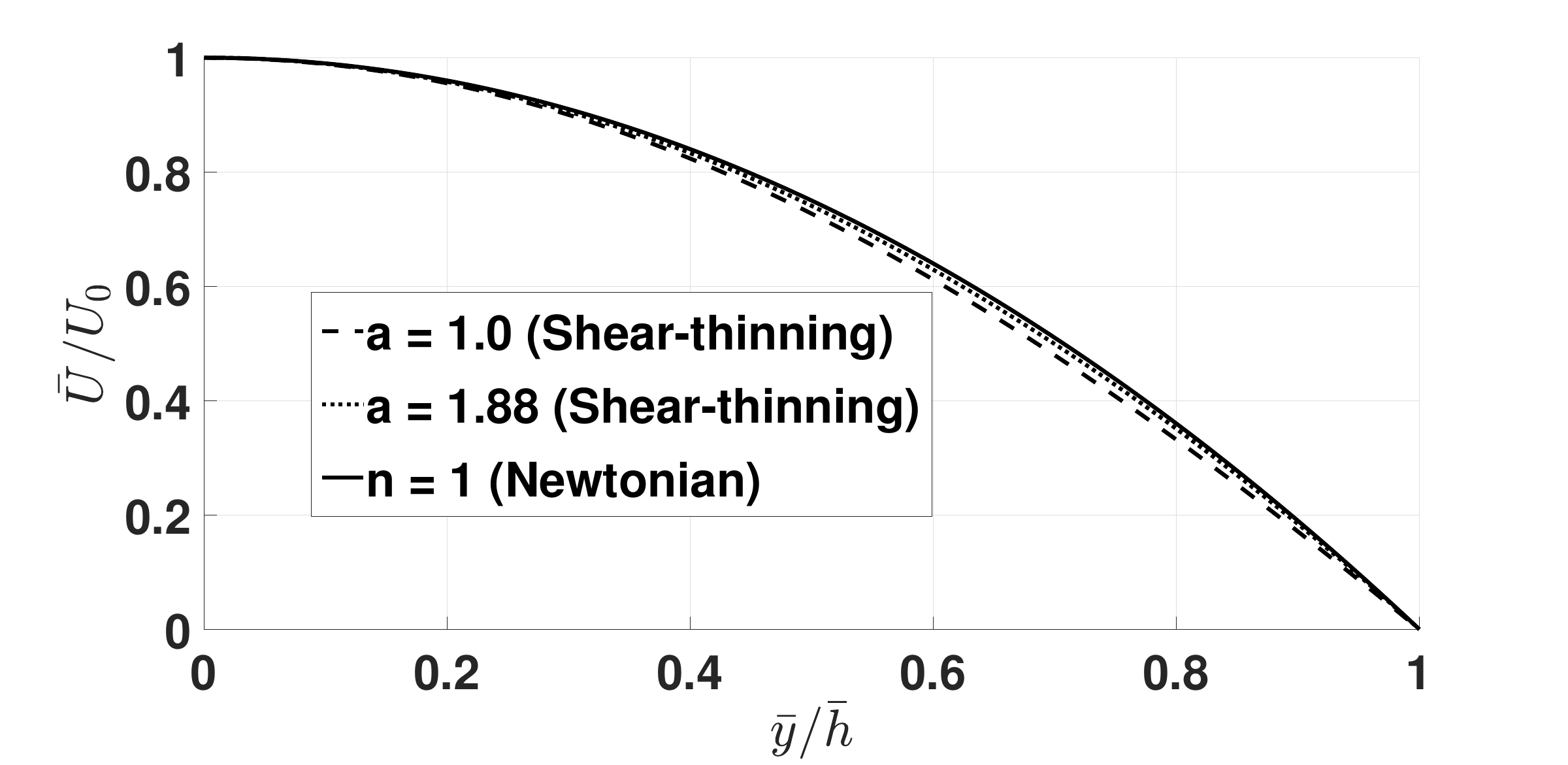}
	\end{center}
	\caption{\textcolor{black}{Normalized velocity $\overline{U}/U_0$ as a function of normalized depth $\overline{y}/\overline{h}$ for two shear-thinning fluids, one with $a$ = 1 and the other with $a$ = 1.88, both having $n$ = 0.5, $I$ = 0 and $L$ = 0.4.}}
	\label{fig12}
\end{figure}

Base state solutions, in terms of film thickness $\overline{h}$ and the surface velocity $\overline{U}(0)$, are shown next for shear-thinning ($n < 1$) and shear-thickening ($n > 1$) fluids. The solutions are given in the physical domain $\overline{y} \in [0;\overline{h}]$. Figures \ref{fig10} and \ref{fig11} show $\overline{h}$ and $\overline{U}(0)$, respectively, for shear-thinning fluids of different intensities, \textcolor{black}{in which we varied 0.2 $\leq$ $n$ $\leq$ 0.6 and 1 $\leq$ $a$ $\leq$ 4 (as $a$ or $n$ tends to zero, shear-thinning effects are stronger)} for fixed $I$ and $L$. The numerical results show a monotonic behavior with both $a$ and $n$, with, as expected, a decrease in the film thickness and an increase in the surface velocity $U_{0}$ (equivalent to a comparison between a high-viscosity shear-thinning fluid and a low-viscosity Newtonian fluid flowing over the same incline). These results are roughly in agreement with the analytical solution of Chimetta and Franklin \cite{chimetta2020analytical}, the exception being a non-monotonic behavior with $a$ in the analytical solution, which can be accounted for by its long-wave approximation (not present in the numerical solution). By fixing $a$ = 2 and the values of $n$, our results are in good agreement with those of Rousset et al. \cite{rousset2007} (see Ref. \cite{chimetta2020analytical} for more details). \textcolor{black}{Figure \ref{fig12} presents the numerical results for the velocity profile considering two shear-thinning fluids, one with $a$ = 1 and the other with $a$ = 1.88, both having $n$ = 0.5, $I$ = 0 and $L$ = 0.4}. We notice that $\dot{\gamma}$ is lower for the shear-thinning fluids, and closer to the wall ($\overline{y}/\overline{h}$ $\rightarrow$ 1) the shear-thinning behavior intensifies.

\begin{figure}[!htb]
	\begin{center}
		\includegraphics[width=0.7\columnwidth]{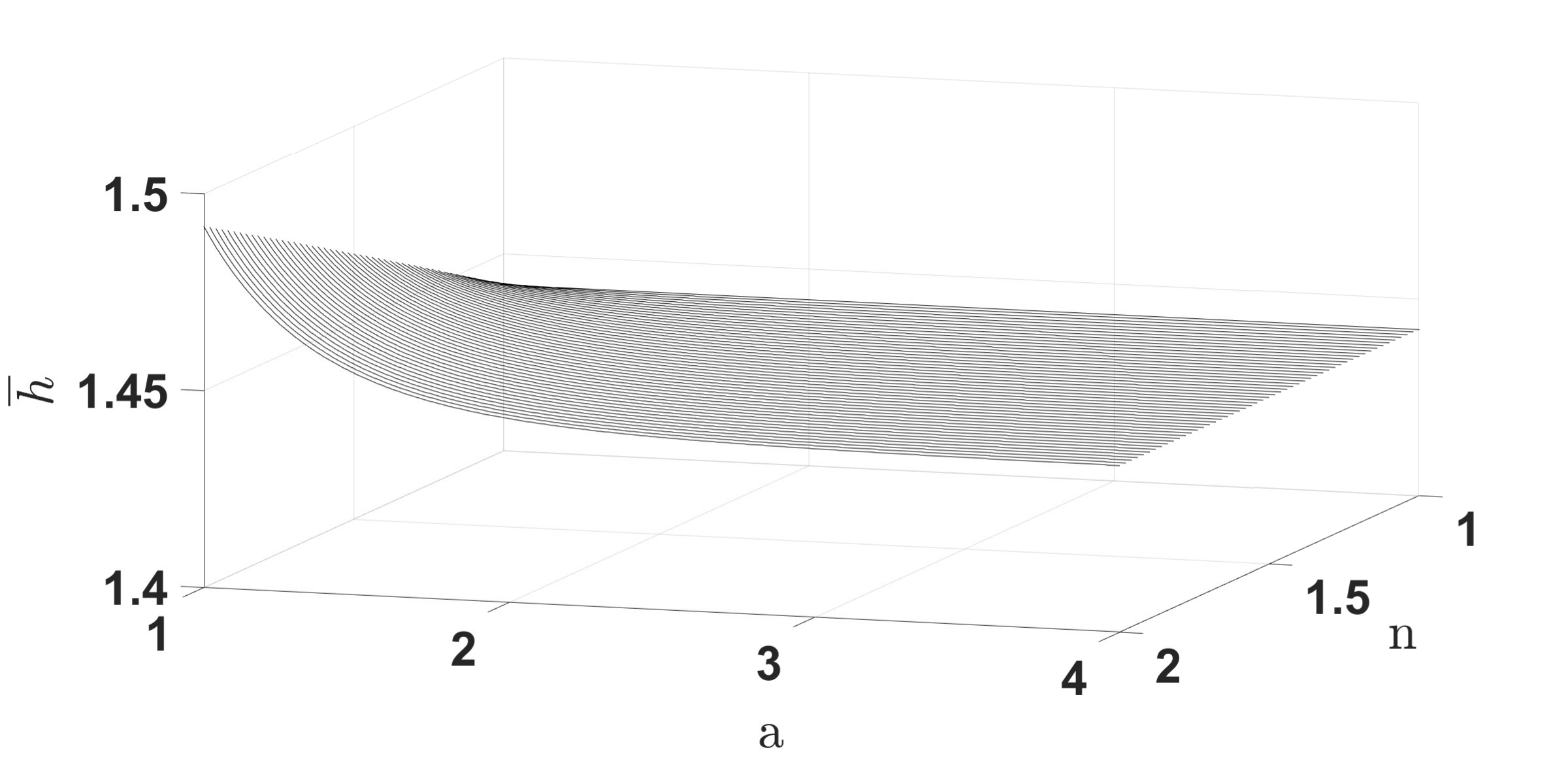}
	\end{center}
	\caption{\textcolor{black}{Numerical result of $\overline{h}$ for shear-thickening fluids (range 1 $\leq$ $n$ $\leq$ 2) with $L$ = 0.4.}}
	\label{fig13}
\end{figure}

\begin{figure}[!htb]
	\begin{center}
		\includegraphics[width=0.7\columnwidth]{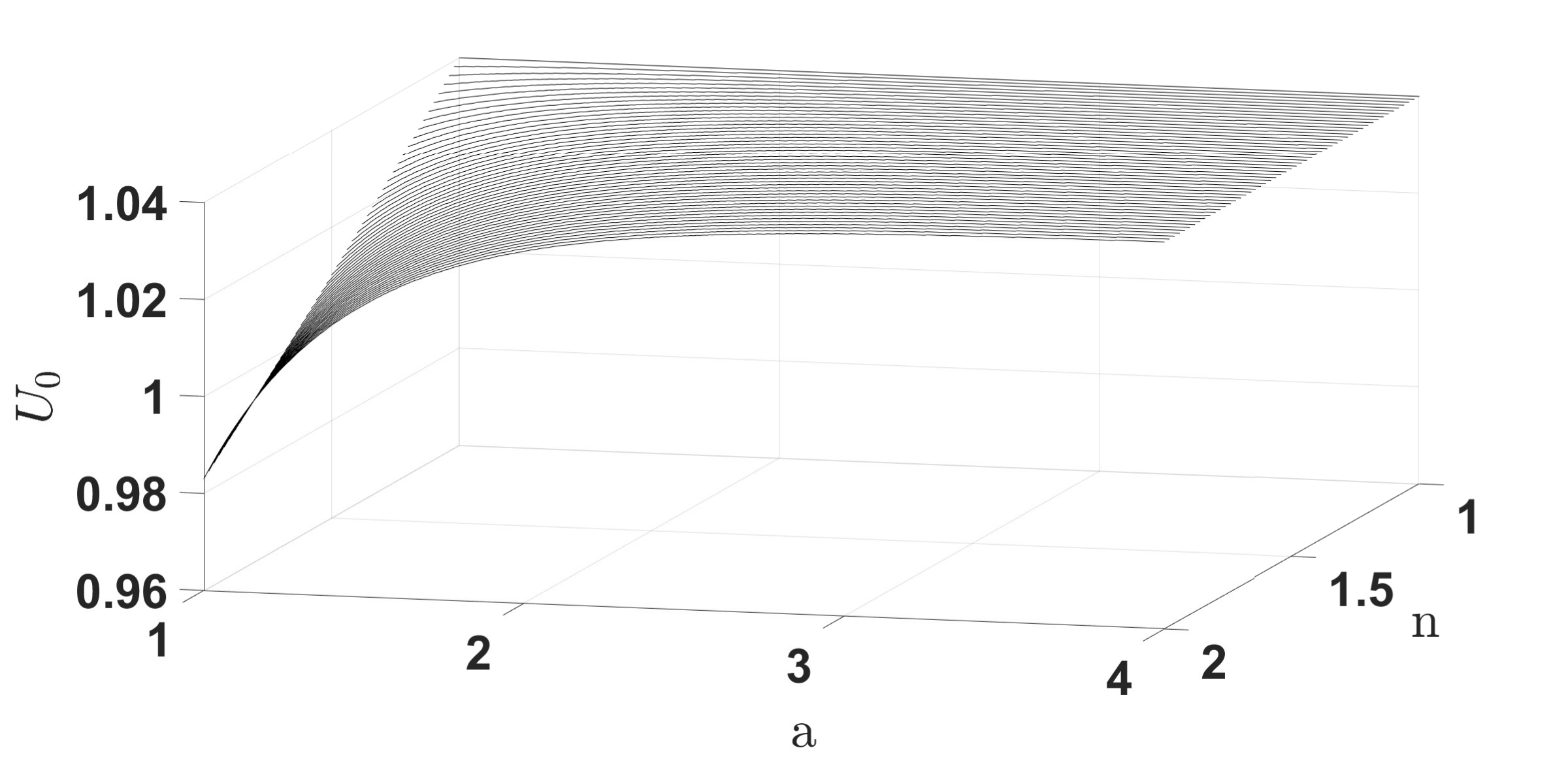}
	\end{center}
	\caption{\textcolor{black}{Numerical result of $U_0$ (surface velocity) for shear-thickening fluids (range 1 $\leq$ $n$ $\leq$ 2) with $L$ = 0.4.}}
	\label{fig14}
\end{figure}

\begin{figure}[!htb]
	\begin{center}
		\includegraphics[width=0.7\columnwidth]{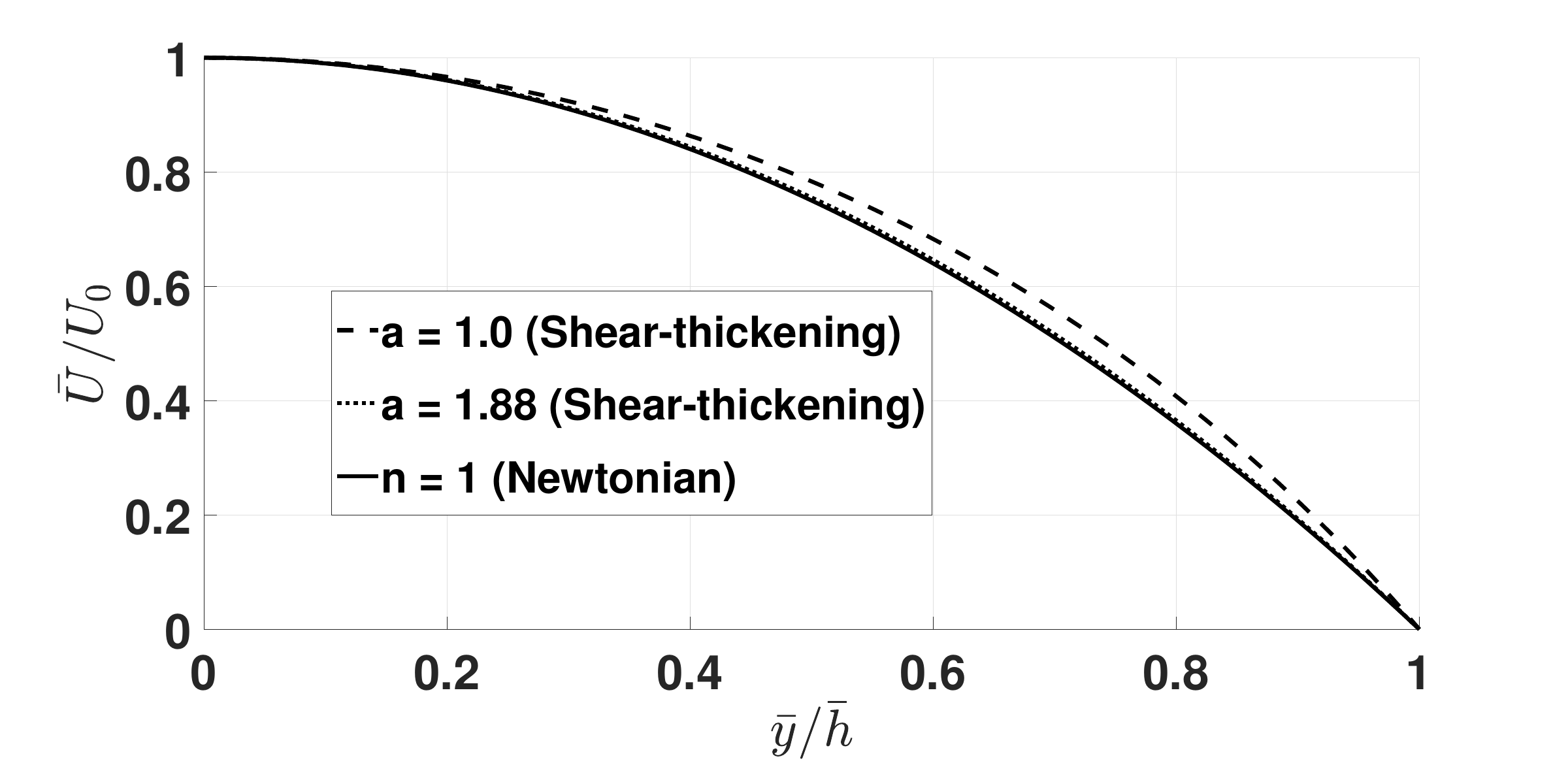}
	\end{center}
	\caption{\textcolor{black}{Normalized velocity $\overline{U}/U_0$ as a function of normalized depth $\overline{y}/\overline{h}$ for two shear-thickening fluids, one with $a$ = 1.0 and the other with $a$ = 1.88, both having $n$ = 2, $I$ = 0 and $L$ = 0.4.}}
	\label{fig15}
\end{figure}

\textcolor{black}{Figures \ref{fig13} and \ref{fig14} show $\overline{h}$ and $U(0)$, respectively, for shear-thickening fluids of different intensities, in which we varied 1 $\leq$ $n$ $\leq$ 2 and 1 $\leq$ $a$ $\leq$ 4 for fixed $I$ and $L$}. As expected, shear-thickening effects are stronger for $a$ $\rightarrow$ 0 or $n$ $\rightarrow$ 2, so that the results are the inverse of those in Figs. \ref{fig10} and \ref{fig11}: greater thicknesses and lower surface velocities as $a$ $\rightarrow$ 0 or $n$ $\rightarrow$ 2. \textcolor{black}{Figure \ref{fig15} presents the numerical results for two shear-thickening fluids, one with $a$ = 1 and the other with $a$ = 1.88, both having $n$ = 2, $I$ = 0 and $L$ = 0.4.} For these cases, we notice that $\dot{\gamma}$ is higher for the shear-thickening fluids. Similarly to Figure \ref{fig12}, as $\overline{y}/\overline{h} \rightarrow 1$ the shear-thickening effects become stronger.

\begin{figure}[!htb]
	\begin{center}
		\includegraphics[width=0.7\columnwidth]{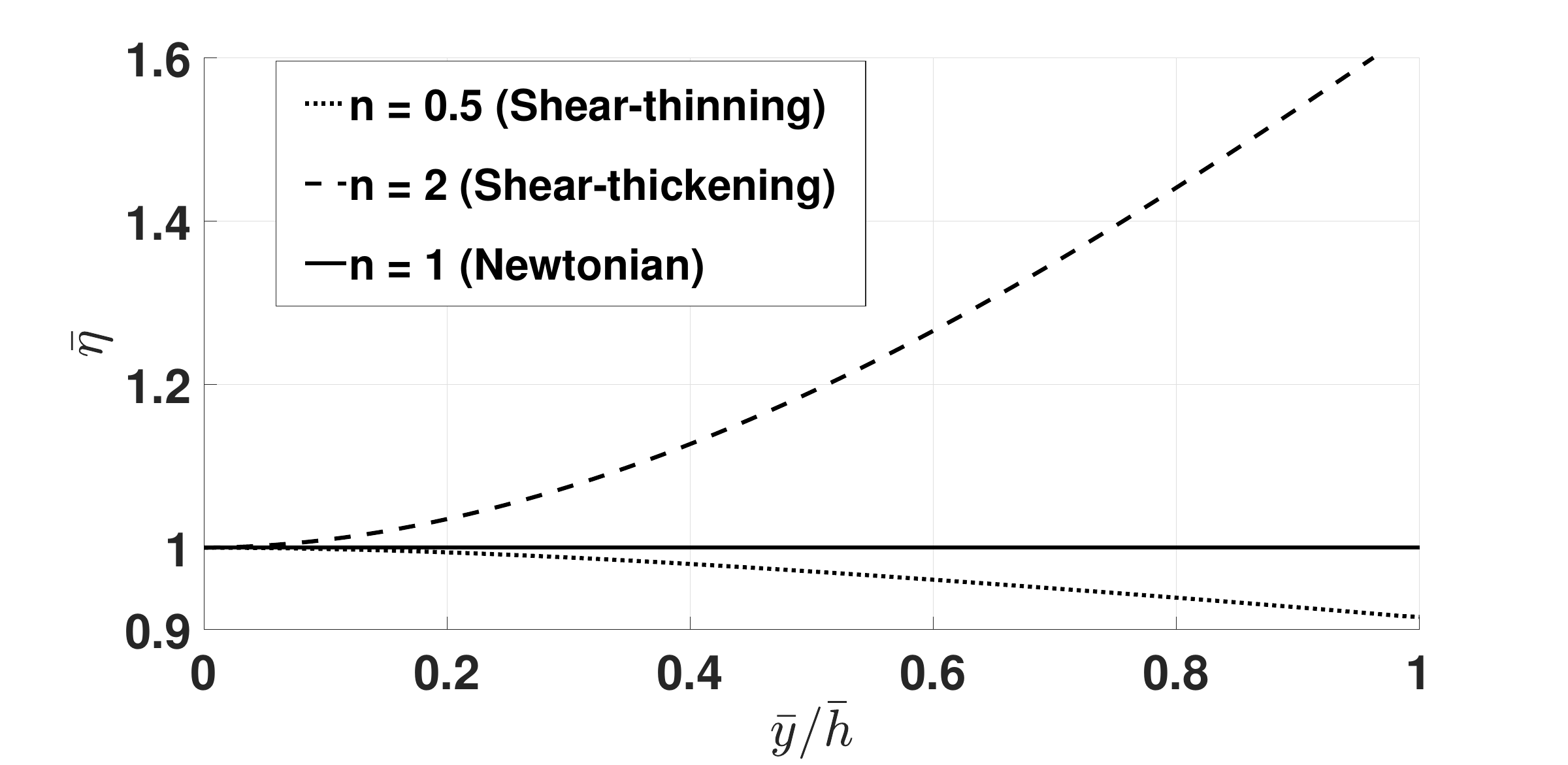}
	\end{center}
	\caption{\textcolor{black}{Viscosity $\overline{\eta}$ as a function of the normalized depth $\overline{y}/\overline{h}$ for shear-thinning ($n$ = 0.5), shear-thickening ($n$ = 2) and Newtonian ($n$ = 1) fluids. In this graphic, $a$ = 1.88, $I$ = 0 and $L$ = 0.4.}}
	\label{fig28}
\end{figure}

Finally, we compare the viscosity of different types of fluids. Figure \ref{fig28} shows the viscosity profiles of \textcolor{black}{shear-thinning ($n$ = 0.5)}, Newtonian ($n$ = 1) and shear-thickening ($n$ = 2) fluids, with $a$ = 1.88, $I$ = 0 and $L$ = 0.4. The profiles corroborate the behaviors found in Figs. \ref{fig12} and \ref{fig15}: closer to the solid surface ($\overline{h}$ $\rightarrow$ 1) non-Newtonian behavior is stronger, while approaching the free surface ($\overline{h}$ $\rightarrow$ 0) all fluids tend to the same behavior.

\subsection{\label{sec3sub@} Perturbations}

We show next the solutions of the Orr-Sommerfeld equation and boundary conditions using the Carreau-Yasuda model (Eqs. \ref{eq38}--\ref{eq44}). Basically, we find $c = c_{r} + c_{i}$, where $c_{r}$ is the wave speed and $c_{i}$ is closely related to the growth rate by $\sigma = \alpha c_{i}$, and the final solutions are given in terms of marginal stability, which define the critical conditions where the film flow is no longer stable (and evolve to a state which eventually present surface waves). Therefore, we solve numerically an eigenvalue problem,  in which $c_{i} = 0$ corresponds to a neutral disturbance, $c_{i} < 0$ to damped disturbances (stable flow), and $c_{i} > 0$ to amplified disturbances (unstable flow). We note that for the $c_{i} > 0$ cases, nonlinear interactions are expected which are not solved in this work (our analysis is linear).

\begin{table}
	\begin{center}
	\caption{\label{tab1}\textcolor{black}{Numerical results for the converged eigenvalue by varying the number of Chebyshev polynomials. All values were obtained with the inverse iteration method implemented with the MATLAB software. Results for a shear-thinning case with $a$ = 1.88, $n$ = 0.5, $I$ = 0, $L$ = 0.4, $\theta = 1^{\circ}$, $We$ = 0.001, $\alpha$ = 0.001 and $Re$ = 1.}}
		\begin{tabular}{cccc}
			&N & $c_{r}$ & $c_{i}$\\
			\hline
			&$5$ & 2.156452697627679 & - 0.062418609481402 \\
			&$10$ & 2.212084543945295 & - 0.064097964113610 \\
			&$20$ & 2.203579709640495 & - 0.064212432580177 \\
			&$30$ & 2.199915786936618 & - 0.064234600029750 \\
			&$40$ & 2.197967881184352 & - 0.064246138921293 \\
			&$50$ & 2.196755762209099 & - 0.064253301018718 \\
			&$60$ & 2.195932367451192 & - 0.064256190418429 \\
			&$70$ & 2.195355140628263 & - 0.064229766157356 \\
		\end{tabular}
	\end{center}
\end{table}

\begin{table}
	\begin{center}
	\caption{\label{tab2}Numerical results for the converged eigenvalue by varying the number of Chebyshev polynomials. All values were obtained with the inverse iteration method implemented with the MATLAB software. Results for a shear-thickening case with $a$ = 1.88, $n$ = 2, $I$ = 0, $L$ = 0.4, $\theta = 1^{\circ}$, $We$ = 0.001, $\alpha$ = 0.001 and $Re$ = 1.}
		\begin{tabular}{cccc}
			&N & $c_{r}$ & $c_{i}$\\
			\hline
			&$5$ & 1.967619702520939 &  - 0.050309073774378 \\
			&$10$ & 1.870876808619691 & - 0.044446802423632 \\
			&$20$ & 1.879047555324375 & - 0.044435069936304 \\
			&$30$ & 1.882413043782644 & - 0.044442893740225 \\
			&$40$ & 1.884214313643472 & - 0.044448022427814 \\
			&$50$ & 1.885281824012552 & - 0.044448385334794 \\
			&$60$ & 1.886095937358268 & - 0.044452669083566 \\
			&$70$ & 1.886627291092958 & - 0.044452804832637 \\
		\end{tabular}
	\end{center}
\end{table}

Tables \ref{tab1} and \ref{tab2} show the values of $c_r$ and $c_i$ for different values of $N$ (number of polynomials) obtained using the function $eig$ as a initial guess for the inverse iteration method (IIM). The parameters used were $a$ = 1.88, $I$ = 0, $L$ = 0.4, $\theta$ = $1^{\circ}$, $We$ = 0.001, $\alpha$ = 0.001, $Re$ = 1 with \textcolor{black}{ $n$ = 0.5 (shear-thinning system)} and $n$ = 2 (for shear-thickening system). Convergence of order $10^{-3}$ for $c_{i}$ is achieved with $N$ = 20 for shear-thinning case, while convergence of order $10^{-4}$ is achieved with $N = 10$ for shear-thickening case. To ensure sufficient accuracy for the next results, all computations were carried out with $N$ = 70.

Before extending further our analysis, we compare the results from our numerical method with those existing in the literature. In particular, the system presented in Rousset et al. \cite{rousset2007} can be reproduced in our model. Therefore, we inserted in our code the parameters used by Rousset et al. \cite{rousset2007}, and present the resulting $Re_c$ in Tabs. \ref{tab3} and \ref{tab4}. These tables do not list the values of $Re_c$ found by Rousset et al. \cite{rousset2007} since the data is presented in graphical form in that paper. Instead, we make reference to Figs. 4 and 6 of that paper. The values of $Re_c$ from our computations show a good agreement with those of Rousset et al. \cite{rousset2007}.

\begin{table}
	\begin{center}
	\caption{\label{tab3}Numerical results for the critical Reynolds number $Re_c$ for different systems. The other parameters were extracted from Fig. 4 of Ref. \cite{rousset2007}.}
		\begin{tabular}{ccccccc}
			& a & n & I & L & $\theta$ & $Re_{c}$ \\
			\hline
			& 2 & 0.5 & 0.00005 & 0 & $10^{\circ}$ & 4.70 \\
			& 2 & 0.5 & 0.00005 & 0.2 & $4^{\circ}$ & 11.35 \\
			& 2 & 0.5 & 0.00005 & 0.4 & $2^{\circ}$ & 20.03 \\
			& 2 & 0.5 & 0.00005 & 0.6 & $1^{\circ}$ & 35.30 \\
			& 2 & 0.5 & 0.00005 & 0.8 & $1^{\circ}$ & 34.47 \\
		\end{tabular}
	\end{center}
\end{table}

\begin{table}
	\begin{center}
	\caption{\label{tab4}Numerical results for the critical Reynolds number $Re_c$ for different systems. The other parameters were extracted from Fig. 6 of Ref. \cite{rousset2007}.}
		\begin{tabular}{ccccccc}
			& a & n & I & L & $\theta$ & $Re_{c}$ \\
			\hline
			& 2 & 0 & 0.00005 & 0 & $1^{\circ}$ & 47.80 \\
			& 2 & 0.6 & 0.00005 & 0.2 & $1^{\circ}$ & 45.80 \\
			& 2 & 0.8 & 0.00005 & 0.4 & $1^{\circ}$ & 44.55 \\
			& 2 & 0.95 & 0.00005 & 0.6 & $1^{\circ}$ & 46.32 \\
			& 2 & 1 & 0.00005 & 0.8 & $1^{\circ}$ & 47.80 \\
		\end{tabular}
	\end{center}
\end{table}

\begin{figure}[!htb]
	\begin{center}
		\includegraphics[width=0.7\columnwidth]{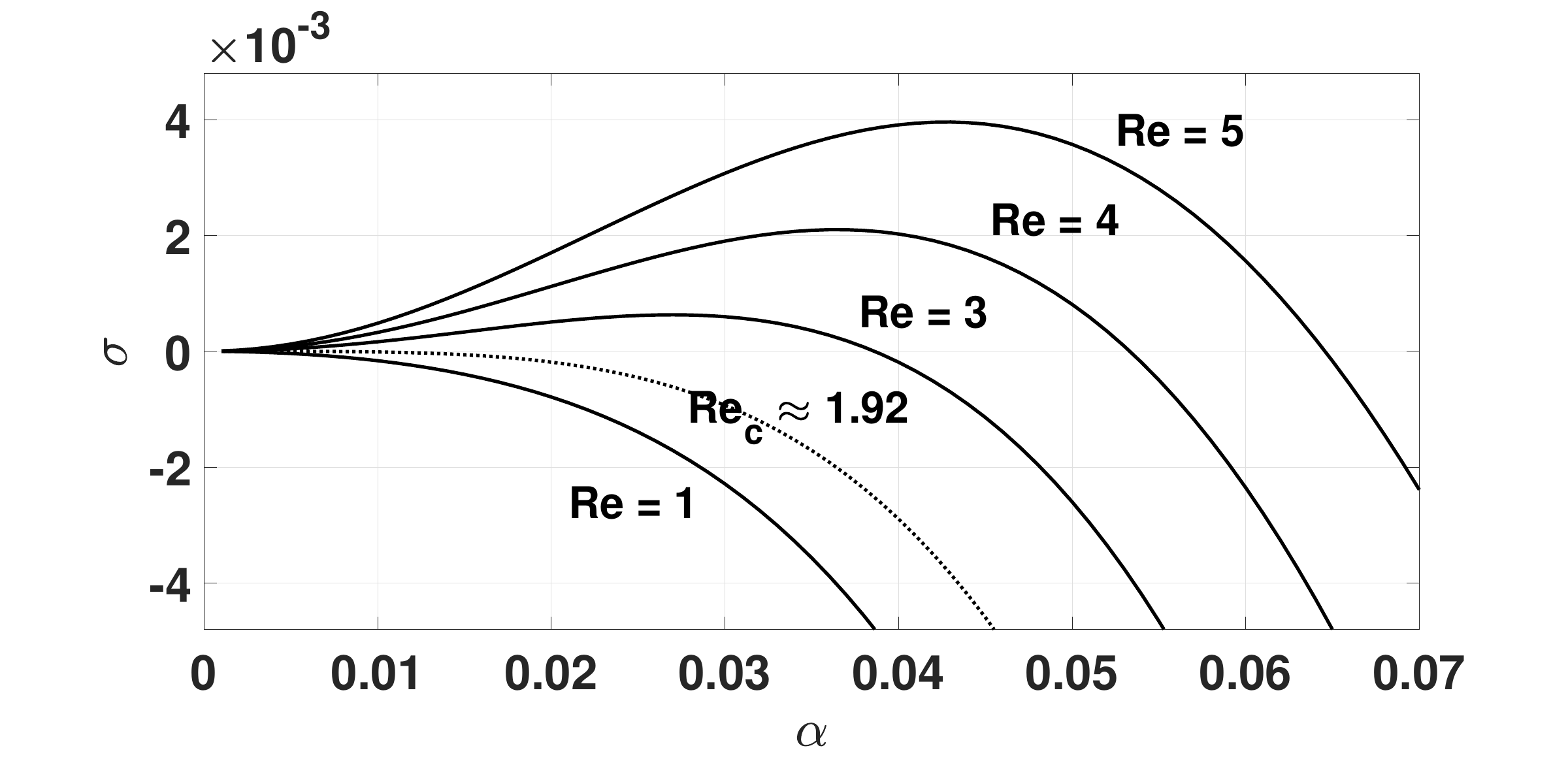}
	\end{center}
	\caption{\textcolor{black}{Dispersion relation $\sigma (\alpha)$ for a shear-thinning fluid, with $a$ = 1.88, $n$ = 0.5, $L$ = 0.4, $\theta = 20^{\circ}$ and $We$ = 0.001.}}
	\label{fig20}
\end{figure}

\begin{figure}[!htb]
	\begin{center}
		\includegraphics[width=0.7\columnwidth]{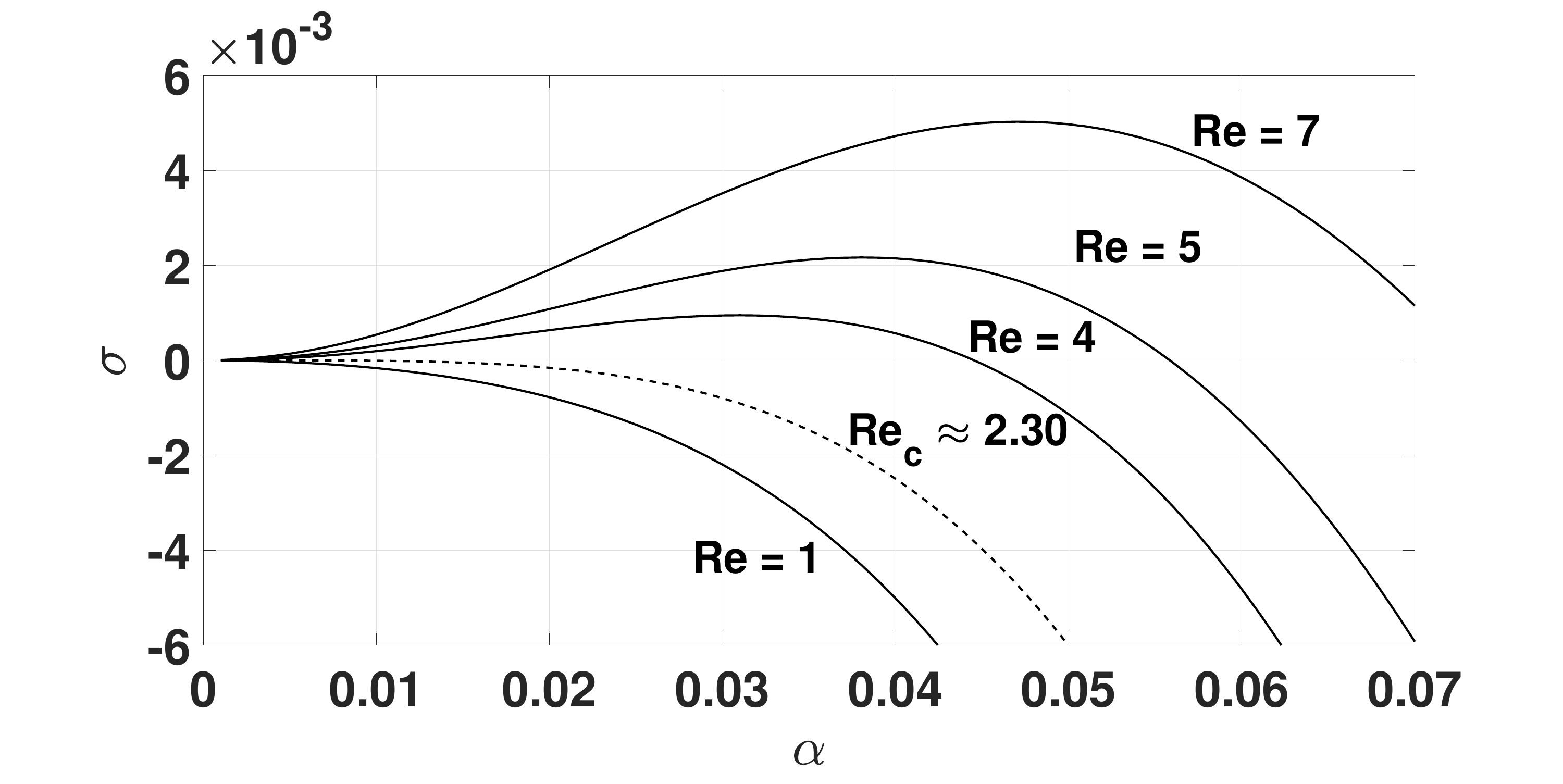}
	\end{center}
	\caption{Dispersion relation $\sigma (\alpha)$ for a Newtonian fluid with $\theta = 20^{\circ}$ and $We$ = 0.001.}
	\label{fig21}
\end{figure}

\begin{figure}[!htb]
	\begin{center}
		\includegraphics[width=0.7\columnwidth]{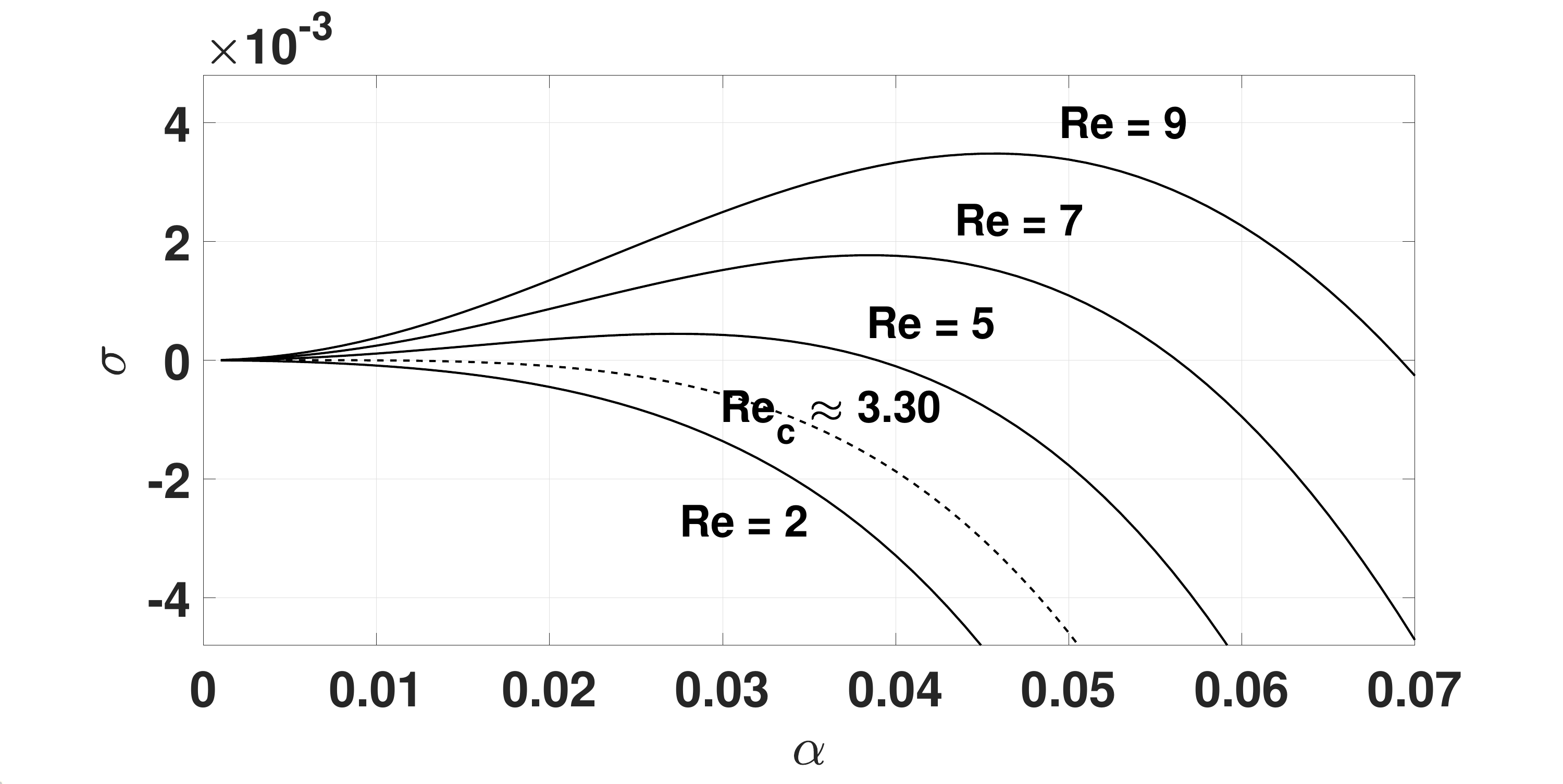}
	\end{center}
	\caption{Dispersion relation $\sigma (\alpha)$ for a shear-thickening fluid, with $a$ = 1.88, $n$ = 2, $L$ = 0.4, $\theta = 20^{\circ}$ and $We$ = 0.001.}
	\label{fig22}
\end{figure}

\begin{figure}[!htb]
	\begin{center}
		\includegraphics[width=0.7\columnwidth]{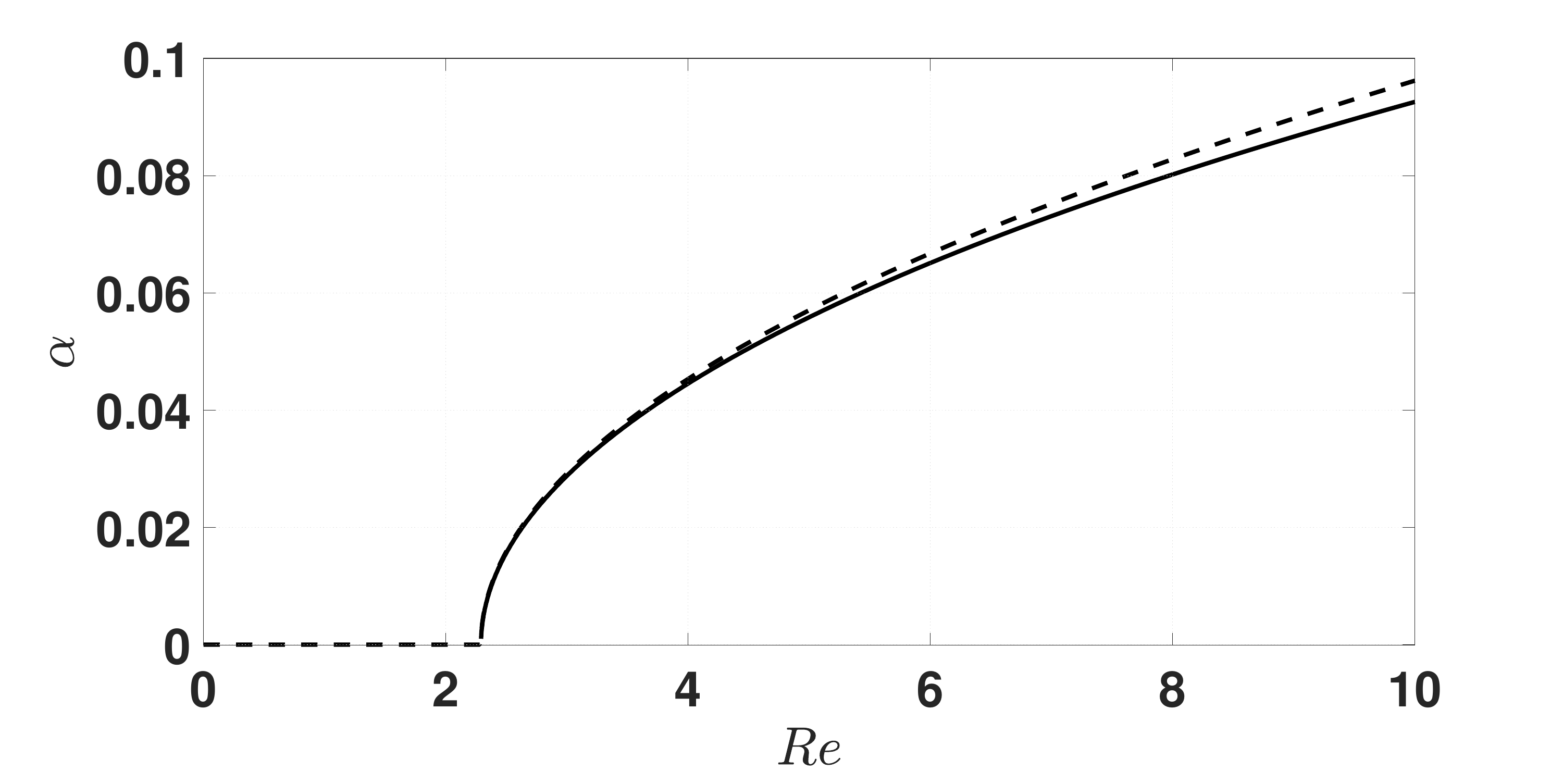}
	\end{center}
	\caption{Neutral stability diagram for the analytical \cite{chimetta2020analytical} and numerical solutions for a Newtonian system with $I$ = 0, $We$ = 0.001 and $\theta = 20^{\circ}$. The continuous curve corresponds to the numerical solution and the dotted one to the asymptotic solution.}
	\label{fig27}
\end{figure}

Figures \ref{fig20}, \ref{fig21} and \ref{fig22} present the growth rate $\sigma$ as a function of the wave number $\alpha$ parameterized by the Reynolds number $Re$ for shear-thinning, Newtonian and shear-thickening fluids, respectively. In these figures, $\theta = 20^{\circ}$ and $We$ = 0.001. By considering that the threshold $\sigma$ = 0 corresponds to the critical conditions for the onset of instabilities, we observe that the critical Reynolds number $Re_{c}$ is approximately 4, 6 and 8 for the shear-thinning, Newtonian and shear-thickening fluids, respectively. Besides, the intervals for $Re_{c}$ show that the shear-thinning fluid presents the lowest value among all cases, with shear-thickening being the highest, and the Newtonian fluid possessing an intermediate behavior value. Therefore, shear-thinning  flows are more susceptible to the emergence of disturbances, since they have higher surface velocities due to low viscosity. Shear-thickening fluids, on the other hand, are more stable because of their higher viscosity values, with lower surface velocities. The Newtonian fluids appears as an intermediate case. These results agree with the expected physical behavior of these flows, since perturbations are attenuated due to the combined effects of viscosity and surface tension. In addition, we compare next our numerical solutions for the Newtonian case with the asymptotic solution of Chimetta and Franklin \cite{chimetta2020analytical}. Figure \ref{fig27} shows the neutral curves for both solutions, showing an excellent agreement. For shear-thinning and shear-thickening fluids direct comparison of both methods are only possible under the assumption of small non-Newtonian effects ($L$ $\rightarrow$ 0), given the limitations of the asymptotic method.

\begin{figure}[!htb]
	\begin{center}
		\includegraphics[width=0.7\columnwidth]{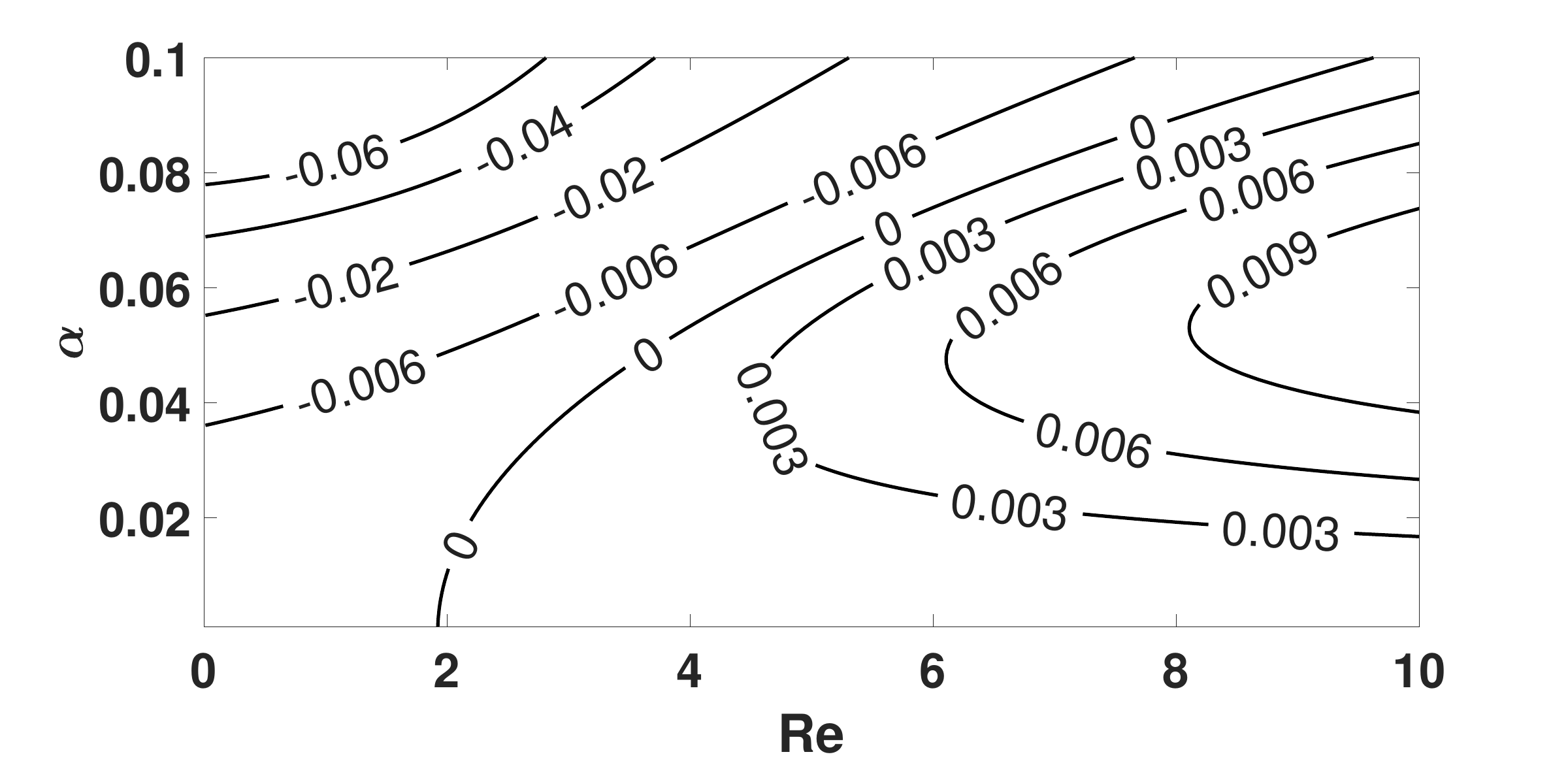}
	\end{center}
	\caption{\textcolor{black}{Neutral stability diagram as a function of Reynolds number with $a$ = 1.88, $n$ = 0.5, $L$ = 0.4, $We$ = 0.001 and $\theta = 20^{\circ}$. Curves with positive and negative $\sigma$ values represent unstable and stable regions, respectively.}}
	\label{fig23}
\end{figure}

\begin{figure}[!htb]
	\begin{center}
		\includegraphics[width=0.7\columnwidth]{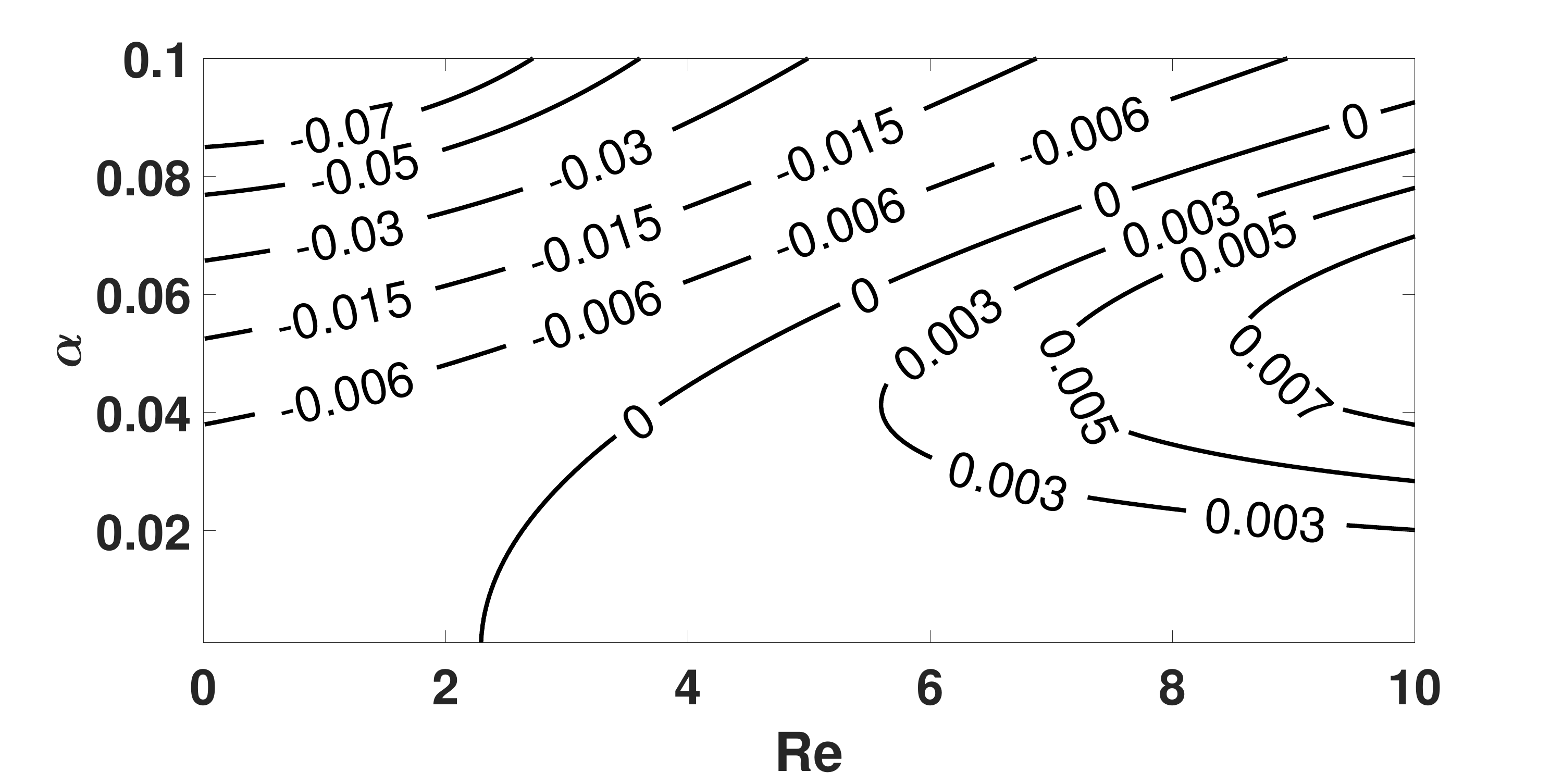}
	\end{center}
	\caption{Neutral stability diagram as a function of Reynolds number for a Newtonian fluid with $We$ = 0.001 and $\theta = 20^{\circ}$.}
	\label{fig24}
\end{figure}

\begin{figure}[!htb]
	\begin{center}
		\includegraphics[width=0.7\columnwidth]{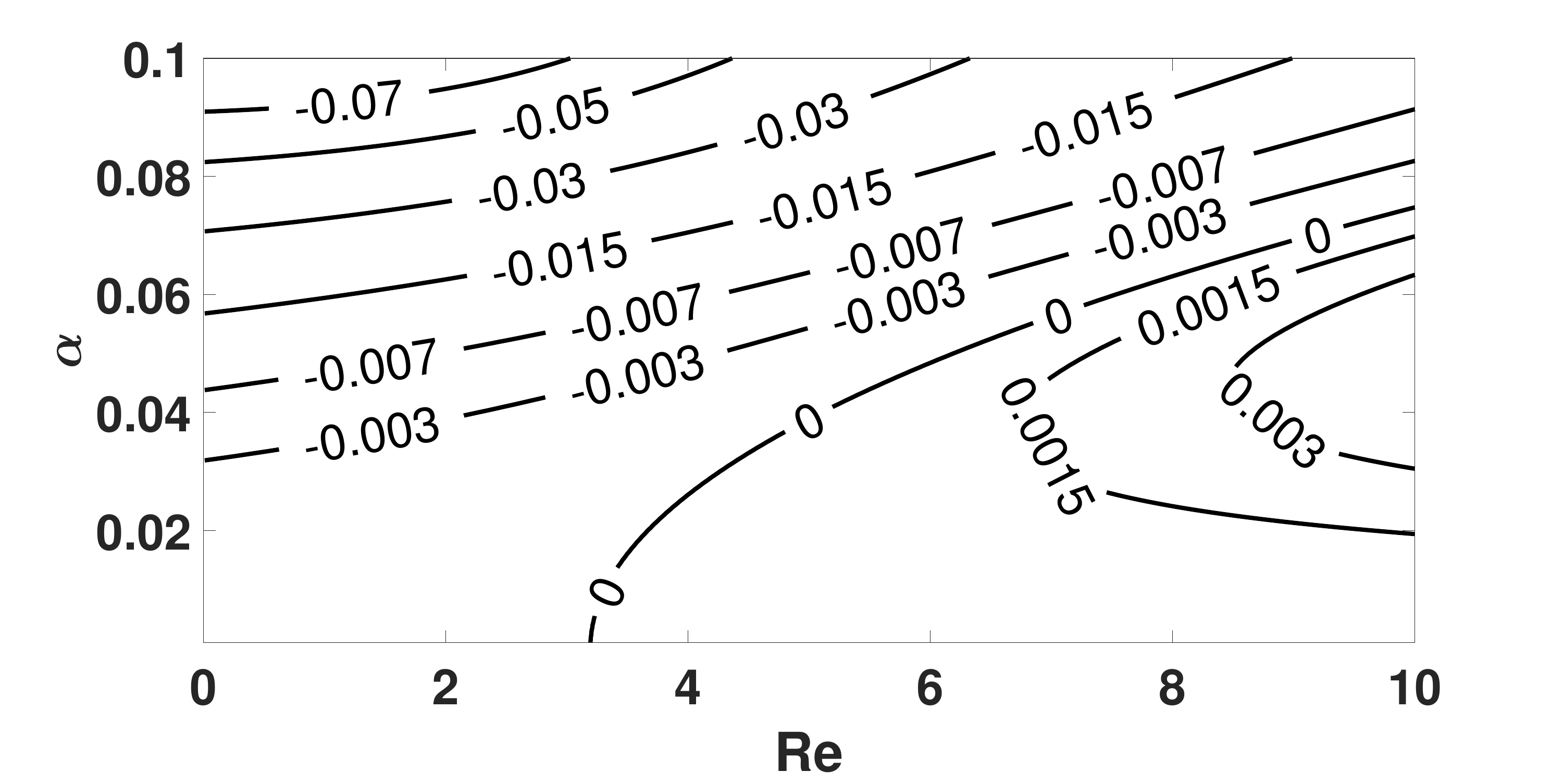}
	\end{center}
	\caption{Neutral stability diagram as a function of Reynolds number with $a$ = 1.88, $n$ = 2, $L$ = 0.4, $We$ = 0.001 and $\theta = 20^{\circ}$.}
	\label{fig25}
\end{figure}

In order to evaluate how the system stability varies with $Re$, we computed neutral stability diagrams in which isocurves of $\sigma$ are plotted as functions of $Re$ and $\alpha$. Figure \ref{fig23} presents the diagram of neutral stability for a shear-thinning fluid with $a$ = 1.88, \textcolor{black}{ $n$ = 0.5,} $I$ = 0, $L$ = 0.4, $We$ = 0.001 and $\theta = 20^{\circ}$. In this diagram, the curve $\sigma$ = 0 represents neutral stability, separating the stable ($\sigma < 0$) and unstable ($\sigma > 0$) bands, and the intersection between the $\sigma$ = 0 and $\alpha = 0$ corresponds to the critical Reynolds number $Re_c$. For this specific shear-thinning case, \textcolor{black}{$Re_{c} = 1.92$} and the unstable band increases with $Re$. Figures \ref{fig24} and \ref{fig25} present the Newtonian ($n$ = 1) and shear-thickening ($n$ = 2) cases, with $Re_{c} = 2.30$ and $Re_{c} = 3.30$, respectively.

\begin{figure}[!htb]
	\begin{center}
		\includegraphics[width=0.7\columnwidth]{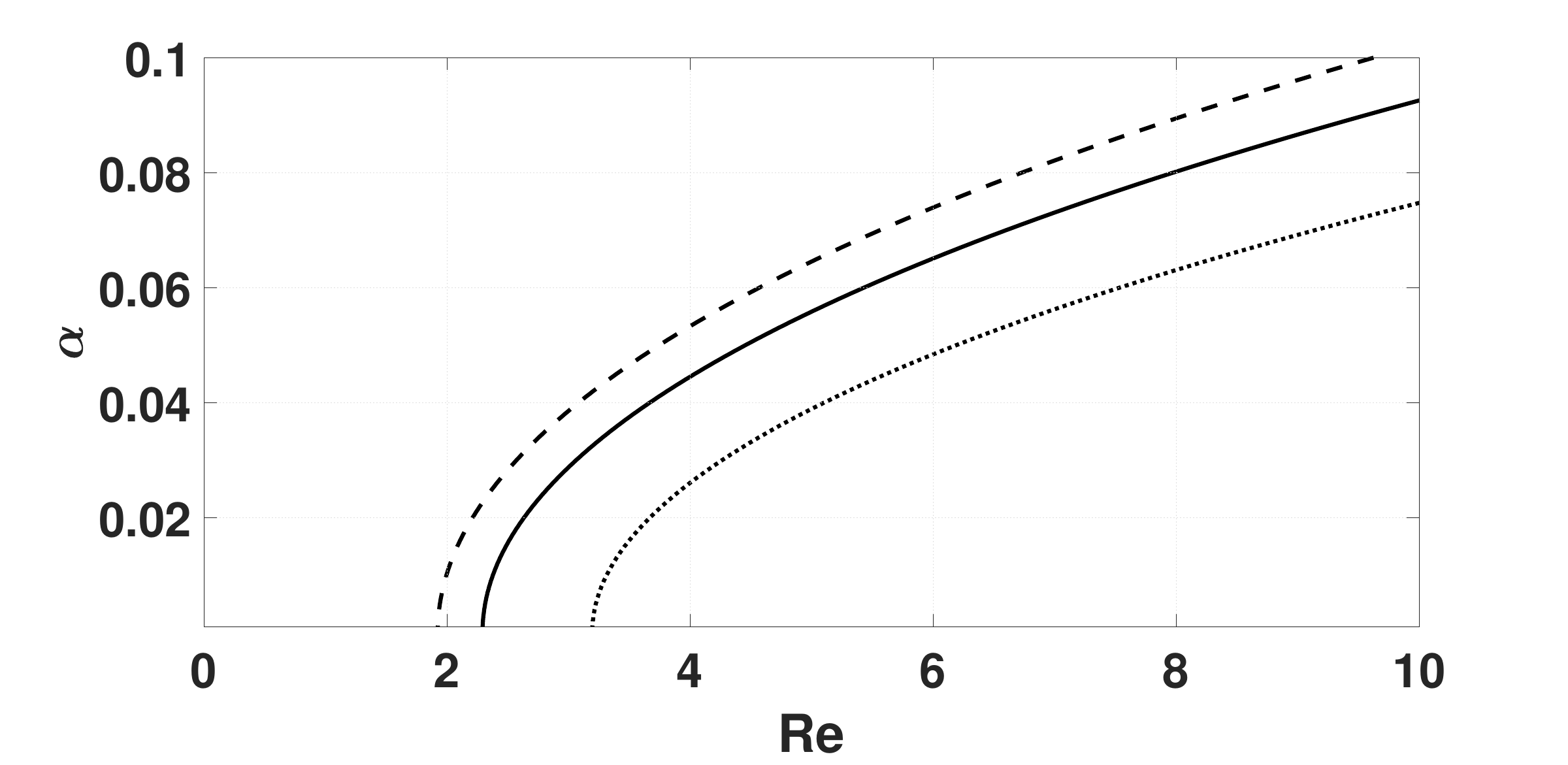}
	\end{center}
	\caption{\textcolor{black}{Diagram of Neutral stability for the shear-thinning ($n$ = 0.5), Newtonian and shear-thickening ($n$ = 2) cases, for $a$ = 1.88, $L$ = 0.4, $\theta = 20^{\circ}$ and $We$ = 0.001. Continuous, dashed and dotted curves correspond to Newtonian, shear-thinning and shear-thickening fluids, respectively.}}
	\label{fig_final}
\end{figure}

For direct comparison between different types of fluid, we plot in Fig. \ref{fig_final} the the marginal stability curves ($\sigma$ = 0) for the shear-thinning, Newtonian and shear-thickening cases. For the three cases the unstable band (to the right of curves) increase with $Re$, \textcolor{black}{and the three critial Reynolds numbers $Re_{c}$ are 1.92, 2.30 and 3.30 for the shear-thinning, Newtonian and shear-thickening flows, respectively.} In other words, the shear-thinning fluid gives the most unstable curve (with the lowest critical Reynolds number), while the shear-thickening case has the highest critical Reynolds number and the smallest unstable band. The Newtonian case appears as an intermediate system between the others. By varying the other parameters, curves $\sigma$ = 0 change. For example, for varying $\theta$, values of $Re_{c}$ change and curves are shifted, but the order of the of cases (more to less stable from shear-thinning to shear-thickening fluids) is preserved. For increasing values of $a$, the curves for the shear-thinning and shear-thickening cases approach the Newtonian intermediate curve, whereas they become farther for decreasing values of $a$ (since non-Newtonian effects are amplified). Finally, a tendency to the Newtonian case occurs for $n \rightarrow 1$ or $L \rightarrow 0$. In summary, the loss of stability implies the growth of surface waves, which are initially two dimensional (but can afterward degenerate into three-dimensional waves, although not investigated in this paper). Shear-thickening fluids are the most stable, that is, the less propense to the appearance of surface waves, while shear-thinning fluids are the most unstable. As the slope is increased, the unstable ranges increase ($Re_{c}$ values become smaller), meaning that surface waves tend to grow for more types of fluids, even shear-thickening fluids.

\section{\label{sec_conclusions} Conclusions}

In this paper, we solved numerically the system studied by Chimetta and Franklin \cite{chimetta2020analytical}, namely the temporal stability of films of non-Newtonian liquids falling by gravity, but without the constrain of long waves. For that, we made use of the Carreau-Yasuda model without assigning specific values to its constants, and proposed a numerical strategy for computing general stability solutions. The numerical strategy is based on expansions of Chebyshev polynomials for discretizing the Orr-Sommerfeld equation and boundary conditions, a Galerkin method for solving the generalized eigenvalue problem, and an Inverse Iteration method to increase accuracy and improve computational time. We ended with a robust and light numerical tool capable of finding the critical conditions for different types of fluids, which we used to analyze Newtonian, shear-thinning and shear-thickening fluids. The numerical outputs can be briefly summarized as: (i) the outputs of the general code match previous solutions obtained for specific computations; (ii) the base state of shear-thinning fluids has higher average velocity and lower thickness, while that of shear-thickening fluids has lower velocity and higher thickness than Newtonian fluids; (iii) for fixed $a$, shear-thinning fluids are the more susceptible to surface instabilities, followed by Newtonian and shear-thickening fluids, the latter being the most stable; (iv) for varying $\theta$, values of the critical Reynolds number $Re_{c}$ change and curves are shifted, but the order of the cases (more to less stable from shear-thinning to shear-thickening fluids) is maintained; (v) for increasing values of $a$, the curves for the shear-thinning and shear-thickening cases approach the Newtonian intermediate curve, whereas they become farther for decreasing values of $a$ (since non-Newtonian effects are amplified); (vi) a tendency to the Newtonian case occurs for $n \rightarrow 1$ or $L \rightarrow 0$; and (vii) the behavior of the neutral stability curve is the same in all cases studied since it is determined by the term $\alpha ^{2} / We$. Our results shed new light on the stability of gravitational flows of non-Newtonian fluids and provide a new tool for obtaining comprehensive solutions on the onset of instability.

\section*{Declaration of Competing Interest}

The authors declare no conflict of interest.

\section*{Acknowledgments}

\begin{sloppypar}
The authors are grateful to the Sao Paulo Research Foundation - FAPESP (Grant No. 2018/14981-7) for the financial support provided.
\end{sloppypar}


\appendix

\section{Weighted residuals method}
\label{appendix_weighted_method}

Weighted residuals methods are defined as approximations in which the residual tends to zero. Consider the scalar product,

\begin{equation}
	\left<f,g\right> _{w} = \int_{a}^{b} f\, g\, \, w\, dx \,\,,
	\label{gal1}
\end{equation}

\noindent where $f(x)$ and $g(x)$ are functions defined on $[a,b]$ and $w(x)$ is a given weight function. It is possible to expand $f(x)$ in a truncated series,

\begin{equation}
	f_{N}(x) = \sum_{k=0}^{N} \hat{f}_{k}\varphi_{k}(x) \ \text{   for   } \ x \in [a;b] \,\,,
	\label{gal2}
\end{equation}

\noindent where $\hat{f}_{k}$ are coefficients to be determined and $\varphi_{k}(x)$ are trial functions. In their turn, the trial functions associated with a given weight $w(x)$ must be orthogonal to be useful in a spectral method,

\begin{equation}
	\left<\varphi_{k}, \varphi_{l}\right>_{w} = c_{k}\, \delta_{k,l} \,\,,
	\label{gal3}
\end{equation}

\noindent where $c_{k}$ is constant and $\delta_{k,l}$ is the Kronecker delta. Now, for a differential equation given by,

\begin{equation}
	L \, f(x) - u = 0 \,\,,
	\label{gal4}
\end{equation}

\noindent it is possible to write the residual as

\begin{equation}
	R_{N}(x) = f(x) - f_{N}(x) \,\,,
	\label{gal5}
\end{equation}

\noindent where $f_{N}(x)$ is the approximate solution and $R_{N}(x)$ is the residual. Combining Eqs. \ref{gal5} and \ref{gal4},

\begin{equation}
	R_{N}(x) = L \, f_{N}(x) - u \,\,,
	\label{gal6}
\end{equation}

\noindent and, by inserting Eqs. \ref{gal2} and \ref{gal4} in Eq. \ref{gal6}, applying a product by the test function $\psi_{i}(x)$ and a weight $w_{\varkappa}$, and integrating over the domain $[a;b]$, one finds

\begin{equation*}
	\int_{a}^{b} R_{N}(x)\psi_{i}(x) w_{\varkappa} d\vec{x} = 
\end{equation*}
\begin{equation*}
	\int_{a}^{b} \Biggl\{ \sum\limits_{k=0}^{N} \hat{f}_{k} L\varphi _{k}(x) - Lf(x) \Biggl\} \psi_{i}(x) w_{\varkappa} d\vec{x} =
\end{equation*}
\begin{equation*}
	\int_{a}^{b} \Biggl\{ \sum\limits_{k=0}^{N} \hat{f}_{k} L\varphi _{k}(x) \psi_{i}(x) w_{\varkappa} \Biggl\} d\vec{x}  - \int_{a}^{b} Lf(x) \psi_{i}(x) w_{\varkappa} d\vec{x} =
\end{equation*}
\begin{equation*}
	\sum\limits_{k=0}^{N} \hat{f}_{k} \Biggl\{ \int_{a}^{b} L\varphi _{k}(x) \psi_{i}(x) w_{\varkappa} d\vec{x} \Biggl\} - \int_{a}^{b} Lf(x) \psi_{i}(x) w_{\varkappa} d\vec{x} \Leftrightarrow
\end{equation*}
\begin{equation*}
	\left<R_{N}(x), \psi_{i}(x)\right> = 
\end{equation*}
\begin{equation}\label{gal7}
	\sum\limits_{k=0}^{N} \hat{f}_{k} \left<L\varphi _{k}(x), \psi_{i}(x)\right> - \left<Lf(x), \psi_{i}(x)\right> \,\,.
\end{equation}

We note that the weight $w_{\varkappa}$ is associated with the trial function and $i \in I_{N}$. Since the method is based on nullifying $R_{N}$ by setting to zero the scalar product $\left<R_{N}(x), \psi_{i}(x)\right>$, the last identity in Eq. \ref{gal7} becomes

\begin{equation}\label{gal8}
	\sum\limits_{k=0}^{N} \hat{f}_{k} \left<L\varphi _{k}(x), \psi_{i}(x)\right> \, = \, \left<Lf(x), \psi_{i}(x)\right> \,\,.
\end{equation}

The Galerkin method is a particular case obtained when the test functions $\psi_{i}$ are chosen from the same family of trial functions $\varphi _{k}$ (therefore, $\psi_{i} = \varphi_{i}$), and the weight $w$ is based on the orthogonality of the trial functions \cite{peyret2013spectral}. Inserting $\hat{f}_{k}$ (obtained by solving Eq. \ref{gal8}) into Eq. \ref{gal2} gives the approximate solution $f_{N}$.

\section{\textcolor{black}{Tangent viscosity}}
\label{tangent_viscosity}

As mentioned in Subsection \ref{sec1sub2}, inserting $\overline{u} = \overline{U} + \hat{u}$ and $\overline{v} = 0 + \hat{v}$ into Eq. \ref{eq5} results in

\begin{equation}\label{pv1}
	\hat{\dot{\gamma}} = \frac{\partial \overline{U}}{\partial \overline{y}} + \frac{\partial \hat{u}}{\partial \overline{y}} + \frac{\partial \hat{v}}{\partial \overline{x}} \,\,,
\end{equation}

\noindent and afterwards, inserting Eq. \ref{pv1} into Eq. \ref{eq4} gives

\begin{equation}\label{pv2}
	\hat{\eta}(\hat{\dot{\gamma}}) = I + (1 - I)\Biggl\{1 + \Biggl[L \Biggl(\frac{\partial \overline{U}}{\partial \overline{y}} + \frac{\partial \hat{u}}{\partial \overline{y}} + \frac{\partial \hat{v}}{\partial \overline{x}}\Biggl)\Biggl]^{a} \Biggl\}^{\frac{n-1}{a}} \,\,.
\end{equation}

\noindent By inserting the perturbations and Eq. \ref{pv2} in Eq. \ref{eq9} we find

\begin{equation}\label{pv3}
	\hat{\tau}_{xx} = 2\Biggl\{I + (1 - I)\Biggl\{1 + \Biggl[L^{a}\Biggl(\frac{\partial \overline{U}}{\partial \overline{y}} + \frac{\partial \hat{u}}{\partial \overline{y}} + \frac{\partial \hat{v}}{\partial \overline{x}}\Biggl)^{a}\Biggl]\Biggl\}^{\frac{n-1}{a}}\Biggl\}\frac{\partial \hat{u}}{\partial \overline{x}} \,\,
\end{equation}

\noindent where, by  using the binomial theorem for $(\frac{\partial \overline{U}}{\partial \overline{y}} + \frac{\partial \hat{u}}{\partial \overline{y}} + \frac{\partial \hat{v}}{\partial \overline{x}})^{a}$ in Eq. \ref{pv3}, we find

\begin{equation}\label{pv4}
	\Biggl(\frac{\partial \overline{U}}{\partial \overline{y}} + \frac{\partial \hat{u}}{\partial \overline{y}} + \frac{\partial \hat{v}}{\partial \overline{x}}\Biggl)^{a} = \Biggl(\frac{\partial \overline{U}}{\partial \overline{y}}\Biggl)^{a} + a\Biggl(\frac{\partial \overline{U}}{\partial \overline{y}}\Biggl)^{a-1}\Biggl(\frac{\partial \hat{u}}{\partial \overline{y}} + \frac{\partial \hat{v}}{\partial \overline{x}}\Biggl) \,\,.
\end{equation}

\noindent Then, inserting Eq. \ref{pv4} in Eq. \ref{pv3} results in

\begin{equation}\label{pv5}
	\hat{\tau}_{xx} = 2\Biggl\{I + (1 - I)\Biggl[1 + \Biggl(L\frac{\partial \overline{U}}{\partial \overline{y}}\Biggl)^{a} + aL^{a}\Biggl(\frac{\partial \overline{U}}{\partial \overline{y}}\Biggl)^{a-1}\Biggl(\frac{\partial \hat{u}}{\partial \overline{y}} + \frac{\partial \hat{v}}{\partial \overline{x}}\Biggl)\Biggl]^{\frac{n-1}{a}}\Biggl\}\frac{\partial \hat{u}}{\partial \overline{x}} \,\,,
\end{equation}

\noindent and, applying the binomial theorem for $[1 + (L\frac{\partial \overline{U}}{\partial \overline{y}})^{a} + aL^{a}(\frac{\partial \overline{U}}{\partial \overline{y}})^{a-1}(\frac{\partial \hat{u}}{\partial \overline{y}} + \frac{\partial \hat{v}}{\partial \overline{x}})]^{\frac{n-1}{a}}$ in Eq.\ref{pv5}, gives

\begin{equation*}
	\Biggl[1 + \Biggl(L\frac{\partial \overline{U}}{\partial \overline{y}}\Biggl)^{a} + aL^{a}\Biggl(\frac{\partial \overline{U}}{\partial \overline{y}}\Biggl)^{a-1}\Biggl(\frac{\partial \hat{u}}{\partial \overline{y}} + \frac{\partial \hat{v}}{\partial \overline{x}}\Biggl)\Biggl]^{\frac{n-1}{a}} = 
\end{equation*}
\begin{equation}\label{pv6}
	\Biggl[1 + \Biggl(L\frac{\partial \overline{U}}{\partial \overline{y}}\Biggl)^{a}\Biggl]^{\frac{n-1}{a}}\Biggl\{1 + (n-1)\Biggl[1 + \Biggl(L\frac{\partial \overline{U}}{\partial \overline{y}}\Biggl)^{a}\Biggl]^{-1}\Biggl(L\frac{\partial \overline{U}}{\partial \overline{y}}\Biggl)^{a}\Biggl(\frac{\partial \overline{U}}{\partial \overline{y}}\Biggl)^{-1}\Biggl(\frac{\partial \hat{u}}{\partial \overline{y}} + \frac{\partial \hat{v}}{\partial \overline{x}}\Biggl)\Biggl\} \,\,.
\end{equation}

\noindent Afterwards, inserting Eq. \ref{pv6} into Eq. \ref{pv5} results in

\begin{equation*}
	\hat{\tau}_{xx} = 2\Biggl\{I + (1 - I)\Biggl[1 + \Biggl(L\frac{\partial \overline{U}}{\partial \overline{y}}\Biggl)^{a}\Biggl]^{\frac{n-1}{a}}\Biggl\}\frac{\partial \hat{u}}{\partial \overline{x}} \Leftrightarrow
\end{equation*}
\begin{equation}\label{pv7}
	\hat{\tau}_{xx} = 2\overline{\eta} \frac{\partial \hat{u}}{\partial \overline{x}} \,\,,
\end{equation}

\noindent where $\overline{\eta}$ is given by,

\begin{equation}\label{pv8}
	\overline{\eta} = I + (1 - I)\Biggl[1 + \Biggl(L\frac{\partial \overline{U}}{\partial \overline{y}}\Biggl)^{a}\Biggl]^{\frac{n-1}{a}} \,\,.
\end{equation}

Following the same procedure for Eq. \ref{eq10} results in

\begin{equation}\label{pv9}
	\hat{\tau}_{yy} = 2\overline{\eta} \frac{\partial \hat{v}}{\partial \overline{y}} \,\,,
\end{equation}

\noindent where $\overline{\eta}$ is given by Eq. \ref{pv8}. Applying the above procedure to Eq. \ref{eq11} gives

\begin{equation}\label{pv10}
	\tau_{xy} = \Biggl\{I + (1 - I)\Biggl\{1 + \Biggl[L \Biggl(\frac{\partial \overline{U}}{\partial \overline{y}} + \frac{\partial \hat{u}}{\partial \overline{y}} + \frac{\partial \hat{v}}{\partial \overline{x}}\Biggl)\Biggl]^{a} \Biggl\}^{\frac{n-1}{a}}  \Biggl\}\Biggl(\frac{\partial \overline{U}}{\partial \overline{y}} + \frac{\partial \hat{u}}{\partial \overline{y}} + \frac{\partial \hat{v}}{\partial \overline{x}}\Biggl) \,\,,
\end{equation}

\noindent which results in

\begin{equation*}
	\tau_{xy} = \Biggl\{I + (1 - I)\Biggl[1 + \Biggl(L\frac{\partial \overline{U}}{\partial \overline{y}}\Biggl)^{a}\Biggl]^{\frac{n-1}{a}}\Biggl\}\frac{\partial \overline{U}}{\partial \overline{y}}
\end{equation*}
\begin{equation*}
	+ \Biggl\{I + (1 - I)\Biggl[1 + n\Biggl(L\frac{\partial \overline{U}}{\partial \overline{y}}\Biggl)^{a}\Biggl] \Biggl[1 + \Biggl(L\frac{\partial \overline{U}}{\partial \overline{y}}\Biggl)^{a}\Biggl]^{\frac{n-1}{a}-1}\Biggl\}\Biggl(\frac{\partial \hat{u}}{\partial \overline{y}} + \frac{\partial \hat{v}}{\partial \overline{x}}\Biggl) =
\end{equation*}
\begin{equation}\label{pv11}
	\overline{\tau}_{xy} + \hat{\tau}_{xy} \,\,,
\end{equation}

\color{black}
\noindent where $\overline{\tau}_{xy} = \overline{\eta}\frac{\partial \overline{U}}{\partial \overline{y}}$, $\hat{\tau}_{xy} = \overline{\epsilon}_{t}(\frac{\partial \hat{u}}{\partial \overline{y}} + \frac{\partial \hat{v}}{\partial \overline{x}})$, and $\overline{\epsilon}_{t}$ is given by

\begin{equation}\label{pv12}
	\overline{\epsilon}_{t} = I + (1 - I)\bigg[1 + n \bigg(L \frac{\partial \overline{U}}{\partial \overline{y}} \bigg)^{a} \bigg] \bigg[1 + \bigg(L \frac{\partial \overline{U}}{\partial \overline{y}} \bigg)^{a} \bigg]^{\frac{n-1}{a}-1} \,\,.
\end{equation}
\color{black}

  \bibliographystyle{elsarticle-num} 
  \bibliography{aipsamp}






\end{document}